\documentclass{geophysics}

\usepackage{amsfonts}
\usepackage{amsmath}
\usepackage{amssymb}
\usepackage{amsxtra}
\usepackage{mathrsfs}

\usepackage{algorithmicx}
\usepackage{algorithm}
\usepackage{algpascal}
\usepackage{graphicx}
\usepackage{mwe}
\usepackage{color}
\usepackage{caption}

\definecolor{darkgreen}{rgb}{0.0, 0.8, 0.1}

\begin{document}

\title{On tomography velocity uncertainty in relation with structural imaging}

\author{
J{\'e}r{\'e}mie Messud, Patrice Guillaume, Gilles Lambar\'e
\\
\textit{CGG, 27 avenue Carnot, 91341 Massy (France)}
}

\righthead{Velocity and migration uncertainties}

\maketitle

\begin{abstract}
Evaluating structural uncertainties associated with seismic imaging and target horizons can be of critical importance for decision-making related to oil and gas exploration and production. 
An important breakthrough for industrial applications has been made with the development of industrial approaches to velocity model building. 
We propose an extension of these approaches, sampling an equi-probable contour of the tomography posterior probability density function (pdf) rather than the full pdf,
and using non-linear slope tomography (rather than standard tomographic migration velocity analysis as in previous publications).
Our approach allows to assess the quality of uncertainty-related assumptions (linearity and Gaussian hypothesis within the Bayesian theory) and estimate volumetric migration positioning uncertainties (a generalization of horizon uncertainties),
in addition to the advantages in terms of efficiency. 
We derive the theoretical concepts underlying this approach and unify our derivations with those of previous publications.
As the method works in the full model space rather than in a preconditioned model space,
we split the analysis into the resolved and unresolved tomography spaces.
We argue that the resolved space uncertainties are to be used in further steps leading to decision-making and can be related to the output of methods that work in a preconditioned model space.
The unresolved space uncertainties represent a qualitative byproduct specific to our method,
strongly highlighting the most uncertain gross areas, thus useful for QCs.
These concepts are demonstrated on a synthetic data.
Complementarily, the industrial viability of the method is illustrated on two different 3D field datasets. The first one shows structural uncertainties on a merge of different seismic surveys in the North Sea. The second one shows the impact of structural uncertainties on gross-rock volume computation.
\end{abstract}

\section*{Introduction}
\label{sec:intro}

Decision-making and risk mitigation are critical for oil and gas exploration and production (E\&P). 
Relying only on maximum-likelihood (or single-valued) subsurface models can lead to drastic misinterpretations of the risk. 
Assessing uncertainties related to maximum-likelihood models is therefore necessary \cite[]{Simpson2000}, but is a challenging task. Indeed, such single-valued models are built by long and complex processes where various types of information are combined sequentially. 
Seismic migration is a central step within those processes,
providing images of the general structure of the subsurface through a reflectivity model.
The positioning uncertainties associated with the structures imaged in the reflectivity,
or structural uncertainties, have been studied for decades \cite[]{Hajnal1981,AlChalabi1994,Thore2002},
in line with advances made in migration tools and workflows.



The key step affecting migration structural uncertainties is velocity model building (VMB). 
According to \cite{Fou13}, VMB is related to one of the biggest ambiguities impacting E\&P. 
While we have seen over the last decade the development of full-wave VMB approaches \cite[]{Virieux2009}, ray-based reflection tomographic approaches remain an essential workhorse method \cite[]{Woodward2008, Guillaume2013,Lambare2014}.
This is due to their inherent characteristics, i.e. efficient numerical implementations, ``compressed'' kinematic data (picks) and ability to digest prior information.
These advantages are particularly appealing from the perspective of a structural uncertainties analysis in an industrial context. 
An important contribution in terms of theory, implementation and application has been delivered by the work of \cite{Osy08a,Osy08b,Osypov2010,Osy11,Osypov2013}. 
While there had been earlier investigations in the context of reflection tomography \cite[]{Duf06},
to our knowledge \cite{Osy08b} were the first to implement a tool used in the industry to estimate structural uncertainties associated with ray-based tomography. 
Their approach is based on the tomography toolbox described by \cite{Woodward2008}. 
The uncertainty analysis is performed around the maximum-likelihood tomography model within a Bayesian approach, assuming  a linearized modeling and a Gaussian probability density function (pdf).
A partial eigen-decomposition of the tomography data operator is performed in a ``model preconditioned basis" \cite[]{Osy08b}. 
This allows the generation of perturbed tomography models related to a confidence level.
Then map (or zero-offset kinematic) migrations of a target horizon within those models give a set of perturbed horizons. 
These are analyzed statistically to derive estimations of horizon error bars related to a confidence level. 
\cite{Osypov2010,Osy11,Osypov2013} give details on practical aspects (such as calibrating the regularizations) and
discuss the impact of anisotropy emphasizing that it represents an important uncertainty component in the seismic images.
They present applications relating to the assessment of oil reserves or well placement,
and demonstrate the affordability and effectiveness of the corresponding horizon uncertainty analysis for industrial applications.

In continuation of these efforts,
recent work on structural uncertainty analysis has been done by
\cite{Messud2017_EAGE, Messud2017_TLE,Messud2018_EAGE}.
The current paper concentrates on the theory underlying this work
and the differences compared to previous works.
The considerations are general and valid for any subsurface model including any form of anisotropy.

The first originality of our approach is, within the Bayesian formalism, to
sample randomly a pdf's equi-probable contour
(related to a clear confidence level) rather than the full pdf \cite[]{Osypov2013,Duffet2002,Duf06},
providing efficient error bar estimates \cite[]{Messud2017_TLE,Messud2017_EAGE,Messud2018_EAGE}.
We define corresponding error bars for a given confidence level,
accounting for the non-diagonal part of the covariance matrices (avoiding some underestimation).

Secondly, while the work of \cite{Osy08a,Osypov2010,Osy11,Osypov2013} was based on the classical (linear) tomographic approach described by \cite{Woodward2008}, our work is based on the non-linear slope tomography of \cite{Guillaume2008} and \cite{Lambare2014}. 
In the context of VMB,
an advantage of non-linear tomography is the ability to compute all non-linear updates of the tomography model with only one picking step \cite[]{Adler2008,Lambare2014},
whereas \cite{Woodward2008} requires a new picking step (thus a new migration) for each iteration (or linear update).
Also, non-linear slope tomography has the advantage of belonging to the family of slope tomography, where the model is recovered from picks of locally coherent reflected events in the pre-stack unmigrated domain \cite[]{Lambare2008}.
In brief, the approach can extract in a non-linear way the kinematic information contained in a dense set of local picks. Numerous versions have been proposed covering a large set of configurations, i.e. multi-layer tomography \cite[]{Guillaume2013b}, dip-constrained tomography \cite[]{Guillaume2013b}, high-definition tomography \cite[]{Guillaume2011}
and joint direct and reflected wave tomography \cite[]{Allemand2017}. 
In the context of structural uncertainty analysis,
one advantage of non-linear tomography together with pdf's equi-probable contour sampling is to provide an efficient way to QC the assumptions made within the Bayesian formalism (linearity and Gaussian pdf hypothesis) \cite[]{Messud2017_EAGE,Reinier2017_EAGE}. 
Another advantage is to make it possible to derive volumetric migration positioning error bars, a volumetric generalization of the horizon positioning error bars that provides migration uncertainties between horizons.

Thirdly, as we work in the full model space rather than in a preconditioned model space, we split the analysis into the resolved and unresolved tomography spaces.
We argue that the resolved space uncertainties are to be used in further steps leading to decision-making and can be related to the output of methods that work in a preconditioned model space.
The unresolved space uncertainties represent a qualitative byproduct specific to our method,
giving an additional information reflects the priors and the illumination,
strongly highlighting the most uncertain areas.
It thus can be used for QCs and may offer some possibility of exploring small-scale non-structural variations, that we cannot consider as fully improbable. 

Note that these works on structural uncertainties can easily apply to full-waveform inversion (FWI)-derived models.
Indeed, these models usually go through a last tomography pass (FWI ``post-processing") in order to obtain flatter common image gathers.
The corresponding tomography uncertainty analysis can then naturally be performed to produce an estimate for FWI model kinematic-related uncertainties.
This workflow is practical as long as rays can adequately describe the kinematics of FWI-derived models. 

The main aim of this paper is to carefully details the theoretical developments associated to our method, providing a unifying framework to compare to other approaches \cite[]{Osy08a,Osypov2010,Osy11,Osypov2013,Duf06}.
Then, the concepts are demonstrated on a synthetic data.
Finally, the industrial viability of the method is illustrated on two different 3D field datasets. The first one shows structural as well as volumetric migration uncertainties on a merge of different seismic surveys in the North Sea. The second one shows the impact of structural uncertainties on gross-rock volume (GRV) computation.

\section*{Tomography and Bayesian formalism}

\subsection*{Tomography inverse problem}
\label{sec:tom}

\begin{figure}[H]
\centering
\includegraphics[width=1.1\linewidth]{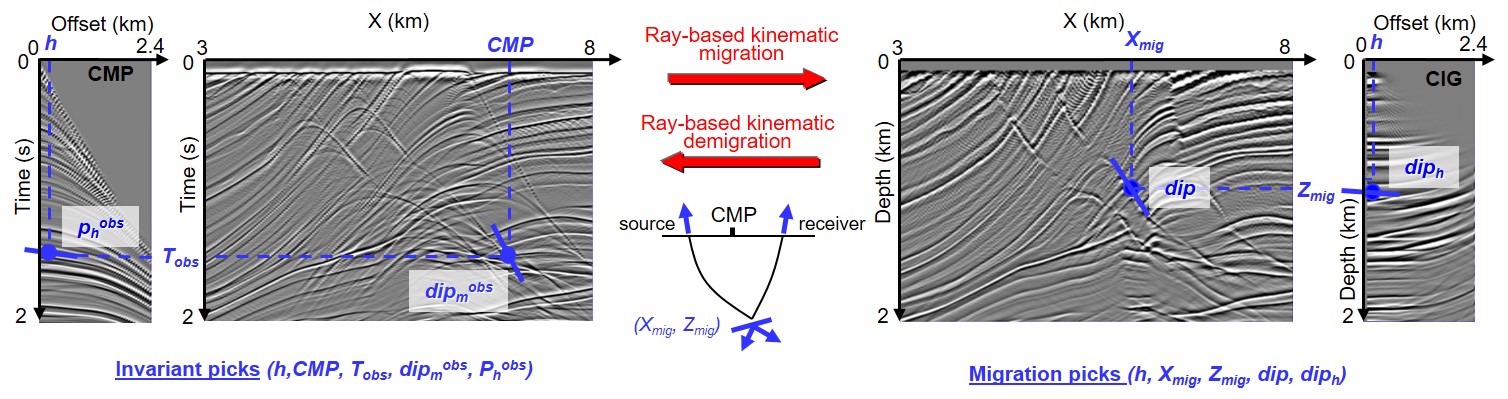}
\caption{
Non-linear slope tomography concepts.
Kinematic invariants represent a set of picks that belong to 
the original data domain.
The invariants can be kinematically migrated in any sufficiently smooth model to deduce corresponding picks in the migrated domain.
Conversely, picks in the migrated domain can be de-migrated in any sufficiently smooth model to deduce corresponding kinematic invariants.
}
\label{fig:figure_tomo}
\end{figure}

Non-linear slope tomography is a practical and efficient tool for velocity model building \cite[]{Guillaume2013}. Its input (or observed) data $\mathbf{d}_{obs}$ consists of a set of ``invariants picks'', i.e. kinematic quantities that belong to the original (multi-dimensional) seismic data domain. Invariants picks are typically described by source and receiver positions, two-way travel-times and two time-slopes describing locally coherent reflection events in the original seismic data domain
(a time-slope in common offset gathers and a time-slope in common mid-point (CMP) gather), see Figure \ref{fig:figure_tomo} (left). 
Invariants picks can be ``kinematically migrated''
by a finite-offset ray-based method in any model $\mathbf{m}$ meeting ray theory assumptions
\cite[]{Duf06,Guillaume2008}, to be converted into locally coherent reflector events in the migrated domain, see Figure \ref{fig:figure_tomo} (right).
Computed events are among others related to two depth-slopes
(a depth-slope in common offset migrated gathers and a depth-slope in common image gather (CIG),
the latter being the local residual move-out (RMO)).
Conversely, locally coherent reflector events in the migrated domain can be picked,
giving ``migration picks'',
and can be ``kinematically demigrated'' in $\mathbf{m}$ by a finite-offset ray-based method, to be converted into invariants picks
\footnote{
The invariant picks can be either directly obtained by picking in the original seismic data domain, or indirectly by a kinematic demigration of picks in the migrated domain. 
The latter is often preferable as the migrated domain has better signal to noise ratio and events separation.
}.

The tomography model $\mathbf{m}$ consists of a set of parameters describing smooth velocity and anisotropy layers; following considerations are thus general and valid for any choice of contribution to $\mathbf{m}$ (velocity and/or any form of anisotropy).
$\mathbf{m}$ is described by cardinal cubic Bspline (ccbs) basis \cite[]{Operto2003}.
Ccbs have many advantages in terms of regularity for the ray tracing (continuous second derivative fields) and computational cost \cite[]{DeBoor1978,Virieux1991}; we will come back to them further.
We denote by $N_M$ the number of nodes that describe $\mathbf{m}$
(500,000 to 50 million in large scale applications).

Non-linear slope tomography finds the model $\mathbf{m}$ that reproduces the best the invariant picks, or equivalently that minimizes the local RMOs \cite[]{Chauris2002}.
The concept of invariant picks allows the implemention of an iterative non-linear inversion scheme where
each linearized update consist of
(1) a non-linear modeling $\mathbf{d}(\mathbf{m})$
(kinematic migration)
and a computation of tomography operator derivatives, and (2) a resolution of an inverse problem to deduce the update.
After each linearized update, the data are modeled again non-linearly
(by kinematic migration)
before a new inversion starts again.
The tomography model $\mathbf{m}$ can be refined by a decomposition into smooth layers (described by ccbs) separated by discontinuities called ``horizons'' (to represent major subsurface contrasts). 
Just like the migrated picks, the horizons describing the layer boundaries can be kinematically demigrated using zero-offset ray tracing to build zero-offset horizon invariant picks.
For each tomography iteration, the horizons are repositioned in the updated model by ``map migration''  (zero-offset ray tracing) of the horizon invariant picks~
\footnote{
This extension is called multi-layer tomography.
Since the process preserves the travel-times, all the layers can be updated simultaneously, avoiding the downward propagation of errors of a conventional top-down approach, and a multi-scale technique avoids getting trapped into a local minimum \cite[]{Guillaume2013}.
}.
The essence of non-linear slope tomography is thus the conversion between invariant picks and migrated picks, that
allows non-linear model updates without having to repick at each iteration.
Priors can easilly be included and fine-tuned, possibly progressively reducing their level to let the data speak more as the number of tomography iterations increases.

This being said, we now formally describe the details of the inversion scheme.
It outputs to the maximum-likelihood tomography model $\mathbf{m}^*$, 
i.e. the model
that fits the best the data under some prior constraints \cite[]{Tarantola1986,Tar05},
defined by
%
\begin{eqnarray}
\mathbf{m}^*
=
\arg\min_{\mathbf{m}}
\frac{1}{2}||{\mathbf{C}}_D^{-1/2}(\mathbf{d}_{obs} - \mathbf{d}(\mathbf{m})) ||_2^2
+
\frac{1}{2}||{\mathbf{C}}_M^{-1/2}(\mathbf{m}-\mathbf{m}_\text{prior}) ||_2^2
.
\label{eq:tomo_pb}
\end{eqnarray}
%
$\mathbf{C}_D$ denotes the covariance matrix in the data space.
It accounts for data (invariant picks) and modeling (kinematic) uncertainties.
$\mathbf{C}_M$ is the ``prior" covariance matrix in the model space,
associated with a prior model $\mathbf{m}_\text{prior}$.
It accounts for uncertainties on the prior model and helps regularization.
Non-zero non-diagonal elements of a covariance matrix $\mathbf{C}_i$ mean that corresponding nodes are correlated.
The diagonal elements of a covariance matrix $\mathbf{C}_i$ are called variances,
and the square roots of the variances are called standard deviations.
%
In the following, bold lowercase letter quantities represent vectors 
and all bold uppercase letter quantities represent matrices.
Some details on our prior and data covariance matrices will be given further.

Equation \ref{eq:tomo_pb} is solved by a non-linear local optimization  method, updating iteratively the model by
\begin{eqnarray}
\mathbf m_{k+1} = \mathbf m_{k} + \delta \mathbf m_{k}
,
\label{eq:m_k+1}
\end{eqnarray}
where $k\in[0,n]$ denotes the iteration number and $n$ the last iteration number.
At the last tomography iteration,
the obtained $\mathbf m_{n}$ is considered to be the maximum-likelihood solution $\mathbf m^*$.
In practice,
the updates $\Delta \mathbf m_{k}$ in equation \ref{eq:m_k+1} are computed through a linearization of equation \ref{eq:tomo_pb}
at each iteration, solving
\begin{eqnarray}
\min_{\delta \mathbf m_{k}}
||
\mathbf{A}_k
\delta \mathbf m_{k}
-
\mathbf{b}_k
||_2^2
,
\label{eq:inv_pb}
\end{eqnarray}
where $\mathbf{A}_k$ is the Jacobian matrix and $\mathbf{b}_k$ the ``error vector", containing information on the data and the prior (these terms are defined in the next paragraphs).
$\mathbf{A}_k$ is defined by
\begin{eqnarray}
&&
\mathbf{A}_k
=
\left[
\begin{array}{c}
\mathbf{C}_D^{-1/2} \mathbf{G}_k \\
\mathbf{C}_R^{-1/2}
\end{array}
\right] 
,
\label{eq:Ak}
\end{eqnarray}
containing the tomography modeling Jacobian matrix at iteration $k$
%
%
\begin{eqnarray}
\mathbf{G}_k
=
\frac{
\partial \mathbf{d}(\mathbf{m})
}{
\partial \mathbf{m}
}
\Big|_{\mathbf{m}=\mathbf{m}_k}
,
\end{eqnarray}
and $\mathbf{C}_R^{1/2}$ gathers various contributions to the prior covariance in model space or constraints
\begin{eqnarray}
&&
\mathbf{C}_R^{-1/2}
=
\left[
\begin{array}{c}
\mathbf{Damp} \\
\mathbf{Other}_1\\
\mathbf{\vdots} \\
\mathbf{Other}_{L-1}
\end{array}
\right]
\nonumber\\
&&
\mathbf{C}_M^{-1}
=
{\mathbf{C}_R^{-1/2}}^\dagger\mathbf{C}_R^{-1/2}
=
\mathbf{Damp}^\dagger\mathbf{Damp}
+ \sum_{i=1}^{L-1} \mathbf{Other}^\dagger_i\mathbf{Other}_i
,
\label{eq:C_R}
\end{eqnarray}
where $^\dagger$ denotes transpose.
Each of the $L$ prior contributions ($\mathbf{Damp}$ and $\mathbf{Other}_i$)
is represented by a square matrix of size $N_M\times N_M$,
$\mathbf{C}_R^{-1/2}$ being a matrix of size $(L\times N_M)\times N_M$.
$\mathbf{Damp}$ represents a ``damping'' diagonal matrix,
scaled for each model contribution or subsurface parameter (velocity, various anisotropies, etc.).
$\mathbf{Other}_i$ represents other possible constraints,
for instance:
\begin{itemize}
\item
A 3D Laplacian to give smoothness to the model layers.
\item
A structural constraint to encourage the model to follow given geological structural dips (parameterized by a curvilinear 2D Laplacian along these dips directions). 
\item
A coupling between pairs of model parameters contributions (velocity Vp, and different anisotropies like $\delta$ and $\epsilon$) to prevent outlier values from rock properties point of view.
A coupling between Vp and $\delta$ can be activated to preserve short spread focusing and a coupling between $\epsilon$ and $\delta$ can be activated to preserve $\eta$.
\item
Possibly a well-ties constraint, etc.
\end{itemize}
Some user-defined parameters enter into these constraints computation,
that are tuned to produce the best tomography result.
For instance, our damping is related to a user-defined ``knob'' tuned so as not to affect the relevant information in the tomography operator,
i.e. to be representative of the tomography noise level.

We emphasize that, in our implementation, all prior contributions to $\mathbf{C}_R^{1/2}$ are $N_M\times N_M$ square matrices.
We do not use non-square matrices $\mathbf{P}$ of size $p\times N_M$ with $p<<N_M$
to parameterize constraints through $\mathbf{P}^\dagger\mathbf{P}$
($\text{rank}(\mathbf{P}^\dagger\mathbf{P})\le p<<N_M$).
In other words, we do not use low-rank prior constraints that reduce
the dimensionality of the problem,
such as steering filters \cite[]{Cla98}.
As already mentioned, our models are described by ccbs grids that
have many advantages in terms of regularity for the ray tracing
and of computational cost \cite[]{DeBoor1978,Virieux1991}.
Dense ccbs grids can be used to represent detailed models up to the maximum resolution that can be obtained from the data. Also, even for large models parameterized by dense ccbs grids, efficient implementations of the tomography inversion can be achieved with current parallel computing capabilities, so that there is no real need for reducing the dimensionality of the problem. Performing the inversion on a dense ccbs grid can have the advantage of not drastically limiting a priori the space of investigated models. With this strategy, the space of investigated models is defined by the ``level of prior'' introduced by the users, selecting the most relevant combination of data and prior information through the user-defined parameters discussed above. 

Denoting by $N_D$ the number of data, i.e. invariant picks, 
the data space covariance matrix $\mathbf{C}_D^{-1/2}$ has size $N_D\times N_D$.
In practice, is often taken diagonal,
consisting of rescaled non-stationary quality weights on the picks
(related among others to the semblances computed by the picking tool).
The error vector has size $(N_D+L\times N_M)$ and is defined by
\begin{eqnarray}
\mathbf{b}_k
=
\left[
\begin{array}{c}
\mathbf{C}_D^{-1/2} ( \mathbf{d}_{obs} - \mathbf{d}(\mathbf{m}_k) ) \\
\mathbf{C}_R^{-1/2} ( \mathbf{m}_k - \mathbf{m}_{prior} )\end{array}
\right] 
.
\label{eq:bk}
\end{eqnarray}
$\mathbf{m}_{prior}$ is often taken to be the model $\mathbf{m}_k$ of the previous iteration, which helps convergence.
This implies that, at the final tomography iteration, the prior tends to be equal to the maximum-likelihood model.

$\mathbf{A}_n$ is a matrix of size $(N_D+L\times N_M)\times N_M$.
The physical dimensions of the components of $\mathbf{A}_k$ are
the inverse of the physical dimensions of the model.
Thus, in the multi-kinds of parameter case (i.e. velocity with anisotropy),
the coefficients of $\mathbf{A}_k$
do not have the same physical dimensions.
A preconditioning can be used:
\begin{eqnarray}
\mathbf{A}_k'
\rightarrow
\mathbf{A}_k\mathbf{D}
,
\end{eqnarray}
where $\mathbf{D}$ is a square matrix that gives the same physical dimension and similar scaling to all coefficients of matrix $\mathbf{A}_k\mathbf{D}$
and can be used to re-weight $\mathbf{A}_n$ to obtain better inversion results.
$\mathbf{D}$ is symmetric, $\mathbf{D}^\dagger=\mathbf{D}$, and should be invertible so that it does not increase the null space of the problem.
A common preconditioning is to choose
$\mathbf{D}$ diagonal and put the inverse of the $L_2$ norm of each column of $\mathbf{A}_k$ on the diagonal.
We then solve, instead of equation \ref{eq:inv_pb},
\begin{eqnarray}
\min_{\delta \mathbf m_{k}'}
||
\mathbf{A}_k'
\delta \mathbf m_{k}'
-
\mathbf{b}_k
||^2
.
\label{eq:inv_pb2}
\end{eqnarray}
and recover $\Delta\mathbf m_{k}$ by
\begin{eqnarray}
\delta \mathbf m_{k}=\mathbf{D}\delta \mathbf m_{k}'
.
\end{eqnarray}

\subsection*{SVD, EVD, LSQR and effective null space projector.}
\label{sec:res_comp}

The solution of equation \ref{eq:inv_pb2}
is usually computed at each iteration using an approximation of
$
\delta \mathbf m_{k}' = \mathbf{A}_k'^{-g}\mathbf{b}_k
$,
where
$
\mathbf{A}_k'^{-g}
=
\lim_{\alpha\rightarrow 0^+}
(
\mathbf{A}_k'^{\dagger}\mathbf{A}_k' + \alpha \mathbf{I}_{N_M}
)^{-1}
\mathbf{A}_k'^{\dagger}
$
is the generalized (or Moore-Penrose pseudo-) inverse,
where $\mathbf{A}_k'^{\dagger}\mathbf{A}_k'$ is called the Hessian matrix in the preconditioned domain.

An approximation of the generalized inverse can be computed for instance performing a partial singular value decomposition (SVD) of $\mathbf{A}_k'$ or a partial eigenvalue value decomposition (EVD) of the Hessian $\mathbf{A}_k'^\dagger\mathbf{A}_k'$.
Suppose that we 
dispose of $p\le N_M$ singular-values $\lambda_{i}$ 
and their corresponding left and right eigenvectors $\mathbf{u}_i$ and $\mathbf{v}_i$
(obtained for instance by $p$ iterations of SVD).
We construct the
$\mathbf{U}_p=[\mathbf{u}_1,, etc.,\mathbf{u}_p]$
and $\mathbf{V}_p=[\mathbf{v}_1,, etc.,\mathbf{v}_p]$ matrices,
and the $p\times p$ matrix $\mathbf{\Lambda}_p^{1/2}$ that contains the
singular values on its diagonal~
\footnote{
$\mathbf{U}_p$ has size $(N_D+L\times N_M)\times p$
and $\mathbf{V}_p$ has size $N_M\times p$.
Note that $\mathbf{U}_p$ and $\mathbf{V}_p$ are not unitary.
They satisfy
$\mathbf{V}_p^\dagger\mathbf{V}_p=\mathbf{U}_p^\dagger\mathbf{U}_p=\mathbf{I}_p
$.
But
$\mathbf{V}_p\mathbf{V}_p^\dagger\ne \mathbf{I}_{N_M}
\text{for $p<N_M$}
$,
and
$
\mathbf{U}_p\mathbf{U}_p^\dagger\ne\mathbf{I}_{N_D+L\times N_M}
$.
}.
The partial SVD of $\mathbf{A}_n'$
gives the best possible rank $p$-approximation of $\mathbf{A}_n'$
\begin{eqnarray}
\mathbf{A}_k'
\approx
\mathbf{U}_p
\mathbf{\Lambda}_p^{1/2}
\mathbf{V}_p^\dagger
\quad\Rightarrow\quad
\mathbf{A}_k'^{-g}
\approx
\mathbf{V}_p
\mathbf{T}_p\mathbf{\Lambda}_p^{-1/2}
\mathbf{U}_p^\dagger
.
\label{eq:A'svd}
\end{eqnarray}
Similarly, the partial EVD of the Hessian is given by
%
\begin{eqnarray}
\mathbf{A}_k'^\dagger\mathbf{A}_k'
\approx
\mathbf{V}_p
\mathbf{\Lambda}_p
\mathbf{V}_p^\dagger
\quad\Rightarrow\quad
(\mathbf{A}_k'^\dagger\mathbf{A}_k')^{-1}
\approx
\mathbf{A}_k'^{-g}\mathbf{A}_k'^{\dagger -g}
\approx
\mathbf{V}_p
\mathbf{T}_p\mathbf{\Lambda}_p^{-1}
\mathbf{V}_p^\dagger
,
\end{eqnarray}
where
\begin{eqnarray}
\mathbf{T}_p
=[ \mathbf{\Lambda}_p + \epsilon \mathbf{I}_p ]^{-1}\mathbf{\Lambda}_p
\label{eq:tikhonov}
\end{eqnarray}
is the Tikhonov regularization operator
that stabilizes the inversion result in the event of very small components of
$\mathbf{\Lambda}_p$
%
\cite[]{Zhang1995,Zha07}.
Note that this regularization is equivalent to including
a ``damping" level $\epsilon$
in a basis where all model components have the same units.
In our case, we take $\epsilon=0$ in $\mathbf{T}_p$
as we already introduced a damping in our prior contributions,
equation \ref{eq:C_R}, that satisfies approximately
$
\mathbf{D}^{-1\dagger}
\mathbf{Damp}^\dagger
\mathbf{Damp}
\mathbf{D}^{-1}
\approx
\epsilon \mathbf{I}_{N_M}
$
in the preconditioned domain.

The $N_M-p$ eigenvectors not resolved by the iterative algorithms
define the tomography ``effective null space",
a cause of multiple equivalent effective solutions
(depending among others on the initial model $\mathbf m_0$) and thus of uncertainty.
The effective null space projector is \cite[]{Mun06}
\begin{eqnarray}
\mathbf{\Pi}_{null_p}
=
\mathbf{I}_{N_M}-\mathbf{V}_p\mathbf{V}_p^\dagger
,
\label{eq:Pp}
\end{eqnarray}
%
%
and the tomography ``resolved" (or partial EVD spanned) space projector is
\footnote{
$\mathbf{R}_{p}\mathbf{\Pi}_{null_p}=\mathbf{\Pi}_{null_p}\mathbf{R}_{p}=\mathbf{I}_{N_M}$.
}
\cite[]{Mun06}
\begin{eqnarray}
\mathbf{R}_{p}
=
\mathbf{V}_p\mathbf{V}_p^\dagger
.
\label{eq:Rp}
\end{eqnarray}

In our implementation, the iterative Least-Squares Quadratic Relaxation (LSQR) 
algorithm is used as a solver for the tomography problem in equation \ref{eq:inv_pb2},
to approximate the effect of $\mathbf{A}_k'^{-g}$ \cite[]{Pai82,Choi06}.
It shares a close similarity with performing an iterative partial SVD or EVD,
but is more efficient as the matrix $\mathbf{A}_k'$ is sparse.
It computes ``Lanczos vectors'', related to the Lanczos tridiagonal matrix, that are not to be confused with eigenvectors \cite[]{Zha07}.
When we wish to recover eigenvectors, necessary for the following uncertainty analysis,
we perform a diagonalization of the Lanczos matrix that gives Ritz vectors,
a sufficient approximation of eigenvectors \cite[]{Zha07}.

\subsection*{Bayesian formalism for tomography model uncertainties}
\label{sec:post_cov}

At the last tomography iteration ($k=n$),
the obtained $\mathbf m_{n}$ is considered to be the maximum-likelihood solution.
The result is uncertain because the tomography input data, modeling, and constraints contain uncertainties.
As a basis for uncertainty considerations,
we will use the Bayesian formalism.
It gives
a clear definition of uncertainties in terms of physics plus a confidence level,
or probability $P$ that the true model belongs to a region of the model space \cite[]{Cow98}.
This is important for reservoir risk analysis.

We consider, within Bayesian theory,
the ``Gaussian posterior pdf in the model space" corresponding to equation \ref{eq:tomo_pb}.
It is defined up to a proportionality constant by \cite[]{Tar05}
\begin{eqnarray}
\tilde\rho_M(\mathbf{m})
\propto
\exp\big[
-\frac{1}{2}
(\mathbf{d}(\mathbf{m})-\mathbf{d}_{obs})^\dagger
{\mathbf{C}}_D^{-1}
(\mathbf{d}(\mathbf{m})-\mathbf{d}_{obs})
-\frac{1}{2}
(\mathbf{m}-\mathbf{m}_\text{prior})^\dagger
{\mathbf{C}}_M^{-1}
(\mathbf{m}-\mathbf{m}_\text{prior})
\big]
.
\label{eq:post_pdf0}
\end{eqnarray}
Take $\mathbf{m}_\text{prior}=\mathbf{m}_n$ in equation \ref{eq:post_pdf0}
(which implies that $\mathbf{m}_\text{prior}$ is chosen to be the previous iteration result,
in agreement with most methods used to solve inverse problems, cf. section ``Tomography inverse problem...").
Consider the first-order (linear) approximation
%
\begin{eqnarray}
\mathbf{d}(\mathbf{m})\approx \mathbf{G}_n (\mathbf{m}-\mathbf{m}_n) + \mathbf{d}(\mathbf{m}_n)
,
\label{eq:lin_g}
\end{eqnarray}
that holds in some region around $\mathbf{m}_n$
(the weaker the non-linearity in $\mathbf{d}(\mathbf{m})$ the larger the region).
Then, equation \ref{eq:post_pdf0} can be rewritten as
%
\begin{eqnarray}
\tilde\rho_M(\mathbf{m})
\propto
\exp\big[
-\frac{1}{2}(\mathbf{m}-\mathbf{m}_n)^\dagger
\tilde{\mathbf{C}}_M^{-1}
(\mathbf{m}-\mathbf{m}_n)
\big]
,
\label{eq:post_pdf}
\end{eqnarray}
where $\tilde{\mathbf{C}}_M$ denotes the ``posterior" covariance matrix in the model space,
defined through
\begin{eqnarray}
\tilde{\mathbf{C}}_M^{-1}
=
\mathbf{G}_n^\dagger \mathbf{C}_D^{-1} \mathbf{G}_n
+
\mathbf{C}_M^{-1}
=
\mathbf{H}_n
.
\label{eq:post_pdf2bis}
\end{eqnarray}
Its inverse $\mathbf{H}_n$ is the posterior Hessian matrix.
%
The maximum-likelihood model $\mathbf{m}_n$ 
does not represent the true model, but the most probable one according to the set of data and priors. Many other probabilistically pertinent models 
(or ``admissible" model perturbations) exist. 
$\tilde\rho_M(\mathbf{m})$ allows a characterization of these models in terms of confidence levels,
giving information on
the confidence region associated with a confidence level $P$ (equal to the integral of $\tilde\rho_M(\mathbf{m})$ over the confidence region).
It thus represents a key to extract uncertainty information with a clear meaning.
Of course, the quality of this interpretation depends on
the approximations of the covariance matrices $\mathbf{C}_D$ and $\mathbf{C}_M$.
In practical applications, this contributes
to the fact that the following uncertainty evaluations will tend to remain somewhat qualitative. This will be further discussed in section ``Global proportionality constraint...''.

Note that $\tilde{\mathbf{C}}_M$, defined through equation (\ref{eq:post_pdf2bis}),
should be computed using the same data and prior covariances than the ones of the (last) tomography pass.
This is assumed in the following.
In other terms, we suppose that all priors for the uncertainty analysis have been defined during the tomography pass (the ones that led to the ``best'' tomography result).
It is of paramount importance that the model is coherent with the assumptions of the uncertainty analysis.

Using notations of section ``Tomography inverse problem...",
the posterior inverse covariance matrix, equation \ref{eq:post_pdf2bis}, can be also computed through
\begin{eqnarray}
\tilde{\mathbf{C}}_M^{-1}
=
\mathbf{A}_n^{\dagger}
\mathbf{A}_n
=
\mathbf{D}^{-1+}
\mathbf{A}_n'^{\dagger}
\mathbf{A}_n'
\mathbf{D}^{-1}
.
\label{eq:A-gA-g+}
\end{eqnarray}
The EVD of $\tilde{\mathbf{C}}_M$,
also defined through the SVD of $\mathbf{A}_n'$,
contains uncertainty information.
Indeed, the principal axes of the Gaussian posterior pdf,
equation \ref{eq:post_pdf},
are given by the eigenvectors of
$\tilde{\mathbf{C}}_M$
and the deformations of the pdf by the eigenvalues of
$\tilde{\mathbf{C}}_M$.
The ``more extended" or poorly resolved directions correspond to smaller eigenvalues of $\tilde{\mathbf{C}}_M$.

\subsection*{Migration structural uncertainties.}
\label{sec:mig_tom}

We are interested in computing 
migration structural uncertainties on target reflectors,
i.e. migration uncertainties related to the kinematic part of the Kirchhoff operator.
Tomography model uncertainties
represent a main contributor to migration kinematic uncertainties.
We can thefore first generate
admissible tomography model perturbations,
i.e. perturbed models
from equations \ref{eq:post_pdf} and \ref{eq:A-gA-g+}.
Then, migrations using each perturbed model can be performed and analyzed.
A study of the results would
allow us to estimate migration kinematic uncertainty-related quantities.

Full Kirchhoff migrations may be considered,
but it is difficult to separate
structural uncertainties from other uncertainties in migrated images,
such as those related to amplitudes.
\cite{Li14} propose to use the Euclidean and Procrustes distances to somewhat perform such a separation.
Here, we are interested mostly in the structural uncertainties related to target reflectors,
a crucial component of migration uncertainties.
We consider a kinematic approximation of the Kirchhoff operator,
the kinematic or map migrations discussed above \cite[]{Duf06,Guillaume2008}.
$\mathbf{h}$ represents the result of the kinematic migration of an invariant pick.
We have
\begin{eqnarray}
\mathbf{h}
=
\mathbf{k}(\mathbf{m})
,
\label{eq:macro_t0}
\end{eqnarray}
where $\mathbf{k}(\mathbf{m})$ is the kinematic migration
operator for an invariant pick,
non linear with respect to the tomography velocity $\mathbf{m}$.
We compute the maximum-likelihood position of a target reflector
$\mathbf{h}_n$ related to the maximum-likelihood tomography model $\mathbf{m}_n$:
\begin{eqnarray}
\mathbf{h}_n
=
\mathbf{k}(\mathbf{m}_n)
.
\label{eq:macro_t1}
\end{eqnarray}
Let us consider a linearization of $\mathbf{k}(\mathbf{m})$ around $\mathbf{m}_n$:
\begin{eqnarray}
(\mathbf{h}-\mathbf{h}_n)
&\approx&
\mathbf{K}^\dagger
(\mathbf{m}-\mathbf{m}_n)
\nonumber\\
\mathbf{K}^\dagger
&=&
\frac{\partial \mathbf{k}(\mathbf{m})}{\partial \mathbf{m}}\Big|_{\mathbf{m}=\mathbf{m}_n}
,
\label{eq:macro_t}
\end{eqnarray}
%
%
where $\mathbf{K}$ represents the linearized approximation 
(or Jacobian matrix) of the kinematic migration operator. 
The migration structural posterior covariance matrix 
related to $\mathbf{h}_n$ is then defined through
%
%
\begin{eqnarray}
\tilde{\mathbf{C}}_K^{-1}
&=&
\mathbf{K}^\dagger
\tilde{\mathbf{C}}_M^{-1}
\mathbf{K}
,
\nonumber\\
(\mathbf{h}-\mathbf{h}_n)^\dagger
\tilde{\mathbf{C}}_K^{-1}
(\mathbf{h}-\mathbf{h}_n)
&\approx&
(\mathbf{m}-\mathbf{m}_n)^\dagger
\tilde{\mathbf{C}}_M^{-1}
(\mathbf{m}-\mathbf{m}_n)
.
\label{eq:macro_t_cov}
\end{eqnarray}
Using notation similar to equation \ref{eq:post_pdf0},
$\tilde{\mathbf{C}}_K$ defines the migration structural posterior pdf $\tilde\rho_K(\mathbf{h})$ in a similar way to equation \ref{eq:post_pdf},
and contains information on migration structural uncertainties (related to the tomography model uncertainties).
%
%

Map migrations (i.e. zero-offset kinematic migrations)
of horizons or target reflectors invariant picks are most often used in practice \cite[]{Duf06,Osy08b,Osypov2013, Messud2017_TLE,Messud2017_EAGE}.
%
Once a set of perturbed models $\{\mathbf{m}\}$
that follow equation \ref{eq:post_pdf} is generated
(and possibly some spurious models removed),
map migrations of horizons or target reflectors may be performed.
This will give a set of perturbed horizons of target reflectors $\{\mathbf{h}\}$ that will
be related to the migration structural posterior pdf according to equation \ref{eq:macro_t_cov}.
Some statistical analysis allows us to deduce structural uncertainty quantities
related to a target reflector and a given confidence level $P$
\cite[]{Duf06,Osy08b,Osypov2013}.

Complementarily and specifically to non-linear slope tomography,
kinematic migrations (using all offsets) of the invariant picks (using all data, not only target reflectors invariant picks) can be considered.
This makes it possible to deduce positioning uncertainty quantities
in the whole migrated volume, not only at target reflector positions.
Details will be given in section ``Specific to non-linear slope tomography...''.

Our analysis accounts for all the important sources of uncertainties
related to the tomography model $\mathbf{m}$, that consists of a set of parameters describing velocity and anisotropy.
Anisotropy, for example, has been pointed out as an important source of uncertainty by \cite{Osy08b,Osypov2010,Osy11}, that is naturally part of the Bayesian formalism presented here. 
Of course, the obtained uncertainties should be interpreted
in the light of the tomographic data that have been used to compute them
(for instance, if faults are not inverted by the tomography,
uncertainties on faults cannot be handled by the method).
Also, the method is valid if :
\begin{itemize}
\item
The tomography model $\mathbf{m}_n$ is a maximum likelihood model, i.e. that lies in the floor of the cost function valley.
Then, perturbations of the model (that stay within the valley) are less probable.
This is assumed in the following.
Even if not fully rigorous, the analysis can also be performed for a model at a local minimum model of the cost function valley, but clearly not for a model on a ``flank" of the valley.
\item
The linearization, equations \ref{eq:post_pdf} and \ref{eq:macro_t}, holds.
The range of validity of the linearization in equation \ref{eq:macro_t} should more or less be the same as that of
the linearization in equation \ref{eq:lin_g}, which allowed us to define $\tilde{\mathbf{C}}_M$ in equation \ref{eq:post_pdf}, that is coherent. This will be further studied below.
\end{itemize}

The main question now is: How to generate
perturbed models
from equation \ref{eq:post_pdf}, together with a confidence level $P$?
We first review previous work in a unifying framework, then highlight questions
and finally present our contribution.

\section*{Sampling a posterior pdf to generate a set of perturbed models}
\label{sec:previous}

\subsection*{Sampling the normal distribution}
\label{sec:gen_th}

Suppose we can find a matrix $\mathbf{E}$
that satisfies
\begin{eqnarray}
\tilde{\mathbf{C}}_M^{-1}
\approx
\mathbf{E}^\dagger
\mathbf{E}
\quad\quad
\Rightarrow
\quad\quad
\tilde{\mathbf{C}}_M
\approx
\mathbf{B}
\mathbf{B}^\dagger
\quad\text{where}\quad
\mathbf{B}=\mathbf{E}^{-g}
.
\label{eq:BB}
\end{eqnarray}
%
%
Then, the posterior pdf, equation \ref{eq:post_pdf}, can be rewritten
\begin{eqnarray}
\tilde\rho_M(\mathbf{m})
\propto
\exp\big[
-\frac{1}{2}\delta\mathbf{r}^\dagger
\delta\mathbf{r}
\big]
\quad\text{where}\quad
\delta\mathbf{r}
=
\mathbf{E}
\Delta\mathbf{m}.
\label{eq:post_pdf2}
\end{eqnarray}
%
$\Delta\mathbf{m}$
represents perturbations of the maximum-likelihood model 
\begin{eqnarray}
\Delta\mathbf{m}=\mathbf{m}-\mathbf{m}_n
.
\label{eq:delta_m}
\end{eqnarray}
Equation \ref{eq:post_pdf2} implies that the covariance matrix associated with the vector
$\delta\mathbf{r}$ is the identity $\mathbf{I}$,
i.e. that the $\delta\mathbf{r}$ coordinates are not correlated and all have a variance of $1$.
One method of generating tomography model perturbations $\Delta\mathbf{m}$ that follow the Gaussian distribution \ref{eq:post_pdf}
is to draw $\delta\mathbf{r}$ from a normal distribution $\mathcal{N}
(\mathbf{0},\mathbf{I})$
and compute
\begin{eqnarray}
\Delta\mathbf{m}=\mathbf{B}\delta\mathbf{r}
,
\label{eq:Br}
\end{eqnarray}
where $\mathbf{B}$ is a 
matrix that
contains the posterior covariance information and scaling.
Once a set of perturbed models $\{\Delta\mathbf{m} + \mathbf{m}_n\}$
is generated (and possibly some spurious models discarded),
migration structural or kinematic uncertainty quantities can be deduced
using the method presented above.

We now discuss possible choices for $\mathbf{B}$ and corresponding dimensionality.
The next three sections describe existing methods and their limitations,
and underline questions still to be clarified.
Then, the rest of the article presents our method.

\subsection*{Cholesky decomposition and partial SVD-based methods}
\label{sec:Duffet}

Suppose we have computed $\tilde{\mathbf{C}}_M^{-1}=\mathbf{A}_n^\dagger\mathbf{A}_n$.
\cite[]{Duf06} propose to perform a Cholesky decomposition
of $\tilde{\mathbf{C}}_M^{-1}$.
This implies finding a lower-triangular square matrix $\mathbf{E}$
of size $N_M\times N_M$ that satisfies (exactly)
equation \ref{eq:BB}
and computing
($\mathbf{E}$ being invertible as $\tilde{\mathbf{C}}_M^{-1}$ is)
\begin{eqnarray}
\mathbf{B}=
\mathbf{E}^{-1}
,
\end{eqnarray}
to generate model perturbations using equation \ref{eq:Br}.
$\mathbf{B}$ is a matrix of size $N_M \times N_M$
and $\delta\mathbf{r}$ is a vector of size $N_M$ draw from $\mathcal{N}(\mathbf{0},\mathbf{I}_{N_M})$.
This scheme is costly
%
%
and may lose accuracy if $\mathbf{A}_n$ is ill-conditioned
\cite[]{Zha07}.


Another method is to use a partial SVD of $\mathbf{A}'_n$, equation \ref{eq:A'svd}.
With equations \ref{eq:A-gA-g+} and \ref{eq:BB},
we deduce
\begin{eqnarray}
\mathbf{E} = 
\mathbf{\Lambda}_p^{1/2}
\mathbf{V}^\dagger_p
\mathbf{D}^{-1}
\quad\quad
\Rightarrow
\quad\quad
\mathbf{B}=\mathbf{E}^{-g}
=
\mathbf{D}\mathbf{V}_p\mathbf{T}_p^{1/2}\mathbf{\Lambda}_p^{-1/2}
,
\label{eq:B_SVD}
\end{eqnarray}
where Tikhonov regularization has been added for the generalized inverse computation.
%
%
$\mathbf{B}$ is a matrix of size $N_M \times p$
and $\delta\mathbf{r}$ a vector of size $p$,
denoted by $\delta\mathbf{p}$ to make the size explicit
and draw from $\mathcal{N}(\mathbf{0},\mathbf{I}_p)$.
This scheme reduces the space of model perturbations,
usually to the degrees of freedom that can be resolved by tomography,
which is numerically advantageous but implies an SVD-based low-rank approximation.
(We did not find applications of this scheme in the literature.)

\subsection*{Prior-based low-rank decomposition method}
\label{sec:P}

This section details the formalism of \cite{Osypov2013,Osy11,Osy08a}
and the specific form obtained for $\mathbf{B}$.
Let us consider prior contributions to the prior covariance matrix in equation (\ref{eq:dec_p}),
that are not represented by square matrices of size $N_M\times N_M$
but by a non-square matrix $\mathbf{P}$ 
of size $N_M\times p$ with $p< N_M$
\begin{eqnarray}
\mathbf{C}_M
\rightarrow
\mathbf{P}
\mathbf{P}^\dagger
.
\label{eq:dec_p}
\end{eqnarray}
As $p<< N_M$ is chosen in practice, low-rank prior constraints are considered,
unlike in section ``Tomography inverse problem..." and equation \ref{eq:C_R}.
The obtained $\mathbf{C}_M$ in equation \ref{eq:dec_p} is not strictly invertible,
i.e. not strictly speaking a covariance matrix
(even if the generalized inverse can be defined).
$\mathbf{P}$ can for instance be a steering filter that contains information on the structures, to reduce
the dimensionality of the problem ($p<<N_M$) \cite[]{Cla98}.
$\mathbf{P}$ is called a preconditioner in \cite{Osypov2013,Osy11,Osy08a},
but it has a very different role from the preconditioner $\mathbf{D}$
introduced in section ``Tomography inverse problem...".
We thus here call it ``model preconditioner".
%

Gathering many constraints within this formalism is done in the model preconditioned basis
(note that $\mathbf{C}_R^{1/2}$ is considered contrarywise to equation \ref{eq:C_R} that considers $\mathbf{C}_R^{-1/2}$)
\begin{eqnarray}
&&
\mathbf{C}_R^{1/2}
=
\left[
\begin{array}{c}
\mathbf{P}\mathbf{Q}_1
\hspace{2mm}
\dots\hspace{2mm}
\mathbf{P}\mathbf{Q}_{L}
\end{array}
\right]
\nonumber\\
&&
\mathbf{C}_M
=
\mathbf{C}_R^{1/2}{\mathbf{C}_R^{1/2}}^\dagger
=
\mathbf{P}
\big( \sum_{i=1}^{L} \mathbf{Q}^\dagger_i\mathbf{Q}_i \big)\mathbf{P}^\dagger
,
\label{eq:C_RP}
\end{eqnarray}
where $\mathbf{Q}_i$ is a $p\times p$ matrix that can for instance contain the
coupling between the various model contributions or subsurface parameters   in the anisotropic case \cite[]{Osypov2013,Osy11,Osy08a}.
To lighten the notations, we do not consider such additional constraints in the following.

We have, using the generalized inverse
\begin{eqnarray}
\tilde{\mathbf{C}}_M
\approx
\big[
\mathbf{G}_n^\dagger \mathbf{C}_D^{-1} \mathbf{G}_n
+
(\mathbf{P}\mathbf{P}^\dagger)^{-g}
\big]^{-g}
=
\mathbf{P}
\big[
(\mathbf{G}_n \mathbf{P})^\dagger \mathbf{C}_D^{-1} (\mathbf{G}_n \mathbf{P})
+
\mathbf{I}_p
\big]^{-1}
\mathbf{P}^\dagger
,
\end{eqnarray}
where
$(\mathbf{G}_n \mathbf{P})^\dagger \mathbf{C}_D^{-1} (\mathbf{G}_n \mathbf{P})+\mathbf{I}_p$
is a $p\times p$ matrix that is invertible
because it is positive definite
($\mathbf{I}_p$ kills the null eigenvalues).
%
Now, let us perform the following partial EVD with $q\le p$ iterations
(no additional preconditioning is needed as the matrices are dimensionless
and the problem is well conditioned in the model preconditioned domain)
%
\begin{eqnarray}
(\mathbf{G}_n \mathbf{P})^\dagger \mathbf{C}_D^{-1} (\mathbf{G}_n \mathbf{P})
\approx
\mathbf{V}_q
\mathbf{\Lambda}_q
\mathbf{V}^\dagger_q
.
\label{eq:EVD_osy}
\end{eqnarray}
%
%
We deduce
\begin{eqnarray}
\tilde{\mathbf{C}}_M
\approx
\mathbf{P}
\big[
\mathbf{V}_q \mathbf{\Lambda}_q \mathbf{V}^\dagger_q
+\mathbf{I}_p
\big]^{-1}
\mathbf{P}^\dagger
.
\end{eqnarray}
%
%
The binomial inverse theorem
recalled in Appendix A leads to
\begin{eqnarray}
\tilde{\mathbf{C}}_M
\approx
\mathbf{P}
\big[
\mathbf{I}_p
-
\mathbf{V}_q \mathbf{V}^\dagger_q
+
\mathbf{V}_q
(\mathbf{\Lambda}_q+\mathbf{I}_q)^{-1}
\mathbf{V}^\dagger_q
\big]
\mathbf{P}^\dagger
.
\end{eqnarray}
Using equation \ref{eq:BB}, \cite{Osypov2013,Osy11,Osy08a} finally obtain 
\begin{eqnarray}
\mathbf{B}
=
\mathbf{P}
\big[
\mathbf{I}_p
-
\mathbf{V}_q \mathbf{V}^\dagger_q
+
\mathbf{V}_q
(\mathbf{\Lambda}_q+\mathbf{I}_q)^{-1/2}
\mathbf{V}^\dagger_q
\big]
,
\end{eqnarray}
allowing to generate perturbations in the model space by equation \ref{eq:Br}.
As here $\delta\mathbf{r}$ is a vector of size $p<<N_M$,
we denote it by $\delta\mathbf{p}$ to make the size explicit.
$\delta\mathbf{p}$ must be drawn from $\mathcal{N}(\mathbf{0},\mathbf{I}_p)$.
$\mathbf{B}$ is a matrix of size $N_M \times p$ that can be split into
\begin{eqnarray}
&&
\mathbf{B}
=
\mathbf{B}^{res}_q
+
\mathbf{B}^{un-res}_{q,p}
\nonumber\\
&&
\mathbf{B}^{res}_q
=
\mathbf{P}
\mathbf{V}_q
(\mathbf{\Lambda}_q+\mathbf{I}_q)^{-1/2}
\mathbf{V}^\dagger_q
\nonumber\\
&&
\mathbf{B}^{un-res}_{q,p}
=
\mathbf{P}
\big[
\mathbf{I}_p
-
\mathbf{V}_q \mathbf{V}^\dagger_q
\big]
\nonumber\\
&&
\quad\quad
\Delta\mathbf{m}^{modprec}=\mathbf{B}\delta\mathbf{p}
\quad\text{where}\quad
\delta\mathbf{p}\sim \mathcal{N}(\mathbf{0},\mathbf{I}_p)
,
\label{eq:Bnullcov}
\end{eqnarray}
where two terms appear:
\begin{itemize}
\item
$\mathbf{B}^{res}_q$ deals with the uncertainty information contained in the posterior covariance matrix and resolved by the partial EVD.
Indeed,
$\mathbf{V}_q (\mathbf{\Lambda}_q+\mathbf{I}_q)^{-1/2} \mathbf{V}^\dagger_q$
represents the partial EVD of $\tilde{\mathbf{C}}_M^{1/2}$ in the model preconditioned domain.
\item
$\mathbf{B}^{un-res}_{q,p}$ contains a projector on an effective null space of dimension $p-q$, as $q$ EVD iterations have been used to approximate a $p\times p$ matrix.
So, only the effective null space of the EVD in the model preconditioned domain
(of size $p$)
can be explored by this method.
Even if this is a very limited part of the full null space of the tomography
operator (because $q\le p<<N_M$),
the advantage is that it to allows to explore some part of the effective null space of the EVD
while keeping geological structures and smoothness
in the model perturbations (the perturbations are projected on $\mathbf{P}$).
\end{itemize}

\subsection*{Questions}
\label{sec:questions}

The methods described in the previous sections have advantages but some questions remain.
In particular, the following points need to be clarified:
\begin{itemize}
\item
A method to QC the validity of the Gaussian and linearized hypothesis.
This is important to check if obtained uncertainties
are pertinent. In the following, we propose an efficient sampling method 
that, together with non-linear slope tomography,
will adress this point.
\item
A link between the methods that work in the full model space and those that work in a preconditioned model space.
We will define the resolved space uncertainty concept that allow to make the connection.
We will also define the unresolved space uncertainties (considering the effective null space projector), that represent a byproduct specific to the following method, giving complementary qualitative information. 
\item
A definition of error bars for the $68.3\%$ (standard-deviation-like) confidence level,
which accounts for the non-diagonal part of the posterior covariance matrices
(as accounting only for the diagonal part would underestimate the errors).
We will give such a definition.
\item
We will also discuss if the computed uncertainties be more than qualitative.
\end{itemize}
%

\section*{A new method sampling a Gaussian equi-probable contour to generate a set of perturbed models}
\label{sec:proposal}

\subsection*{Sampling a Gaussian equi-probable contour}
\label{sec:equip}

Let us return to the tomography posterior Gaussian pdf, equation \ref{eq:post_pdf},
and propose a different sampling method that will reduce the sampled space and thus optimize the exploration.
We consider an equi-probable contour
\begin{eqnarray}
\mathbf{\Delta m}^\dagger
\tilde{\mathbf{C}}_M^{-1}
\mathbf{\Delta m}
=
Q_{N_M}(P)
,
\label{eq:iso_pdf}
\end{eqnarray}
where
$Q_{N_M}(P)$ 
is the quantile of order $P$ of the Chi-square distribution \cite[]{Cow98}
(this distribution is among others used in confidence interval estimations related to 
Gaussian random variables).
%

\begin{figure}[H]
\centering
\includegraphics[width=0.6\linewidth]{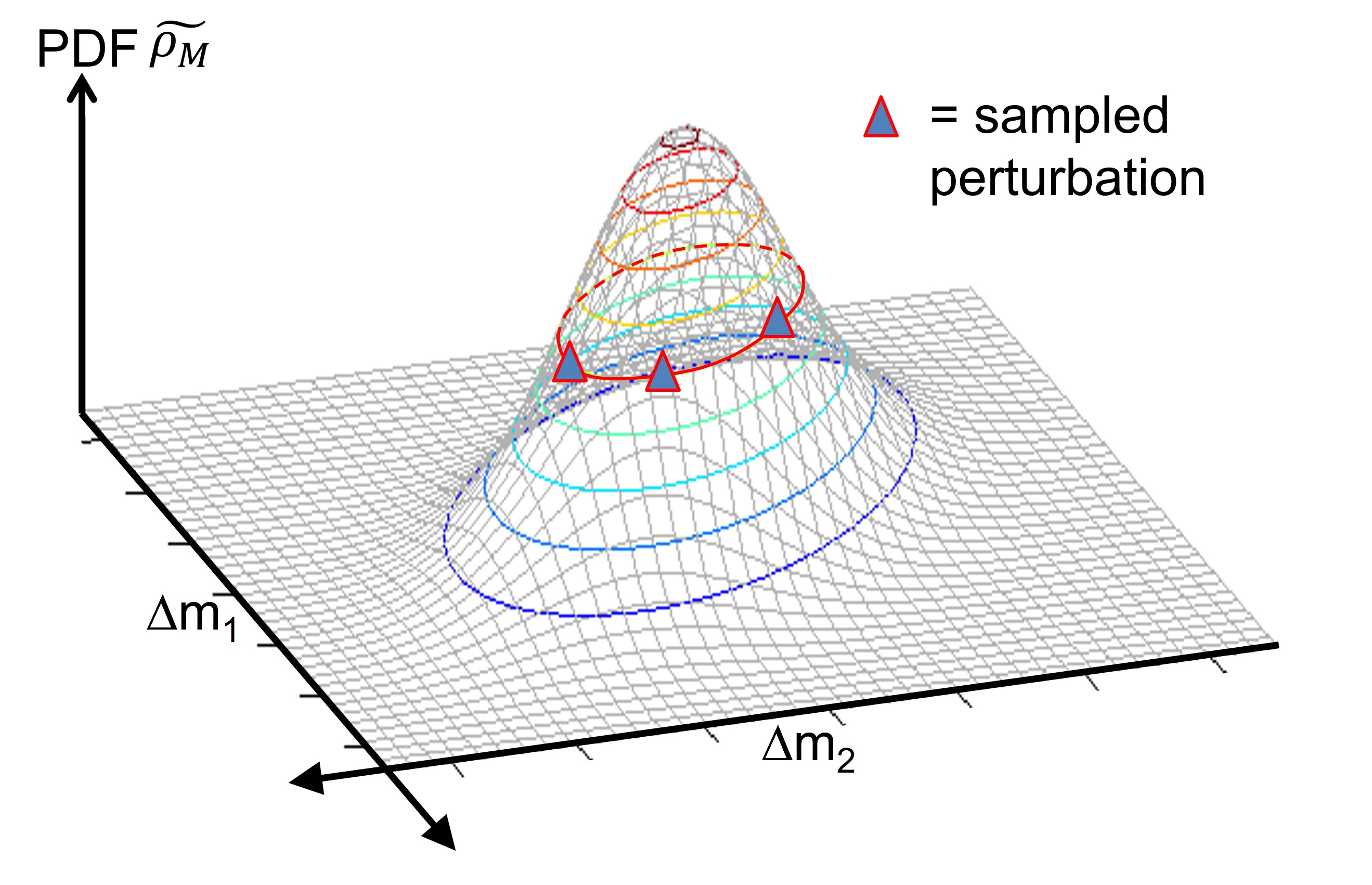}
\caption{
Equi-probable contour of a posterior Gaussian pdf  ($N_M=2$) (red)
and $\Delta\mathbf{m}$ samples (triangles).
}
\label{fig:figure1}
\end{figure}

Resolving (or sampling) equation \ref{eq:iso_pdf} for a given $P$ value
gives the set of maximum perturbations (or the boundary of the confidence region)
associated with a confidence level $P$;
Figure \ref{fig:figure1} gives an illustration.
The probability that the true model $\mathbf{m}$
lies within the $N_M$-dimensional hyper-ellipsoid
%
%
of center $\mathbf{m}_n$
defined by equation \ref{eq:iso_pdf}
is equal to $P$.
Restricting the sampled space to an equi-probable contour does not hamper the assessment of the uncertainties compared to the sampling of the full pdf because information on the full Gaussian is contained in one contour and the corresponding $P$ value.
Indeed, all hyper-ellipsoids defined by equation \ref{eq:iso_pdf} are related by a simple proportionality constant. 

In the following, we consider a confidence level $P=68.3\%$.
We call it the standard-deviation-like confidence level \cite[]{Messud2017_EAGE}
because it corresponds to a standard deviation interval when $N_M=1$ or when model parameters are non-correlated and their single pdfs are considered independently, see Appendix B. 
For more generality, we define uncertainties through the tomography confidence region related to a probability $P=68.3\%$
and resolve equation \ref{eq:iso_pdf} to generate a set of admissible perturbations.
In the spirit of the considerations of section ``Sampling a normal distribution...",
the solution is
\begin{eqnarray}
\mathbf{\delta r}^\dagger\mathbf{\delta r}=Q_{N_M}(68.3\%)
\quad\text{and}\quad
\tilde{\mathbf{C}}_M
\approx
\mathbf{B}
\mathbf{B}^\dagger
\quad\Rightarrow\quad
\Delta\mathbf{m}
=
\mathbf{B}\delta\mathbf{r}
,
\label{eq:sol_hyp_elips}
\end{eqnarray}
where $\mathbf{\delta r}$ is a vector of size $N_M$ drawn from a uniform distribution and rescaled to have norm $||\mathbf{\delta r}||_2=\sqrt{Q_{N_M}(68.3\%)}$
(it must not be drawn from a Gaussian distribution here, 
as we sample only an equi-probable contour).
%
Generating model perturbations from an equi-probable contour sampling
has the advantage of reducing the sampled space
to its most representative components,
optimizing the computation.
In our case, where we do not use low-rank prior constraints that reduce
the dimensionality of the problem ($\mathbf{B}$ is a $N_M\times N_M$ matrix),
the method allows us to obtain stable uncertainty estimates with 200-500 random models, for large-scale applications.
And especially, as detailed further, 
pdf's equi-probable contour sampling together with non-linear tomography provide an efficient way to QC the assumptions made within the Bayesian formalism (linearity and Gaussian pdf hypothesis).

Let us now specify a problem concerning error bars.
Why not simply define error bars by
$\pm\sqrt{\text{diag} (\tilde{\mathbf{C}}_M)}$,
where $\text{diag} (\tilde{\mathbf{C}}_M)$
denotes the vector containing the diagonal elements of $\tilde{\mathbf{C}}_M$?
This is sometimes used as a first tomography model uncertainty indicator \cite[]{Osypov2013}, but it has pathologies.
%
When $\tilde{\mathbf{C}}_M$ is diagonal,
a subset of the solutions of equation \ref{eq:iso_pdf} with $P=68.3\%$ is
%
%
\begin{eqnarray}
\Delta\mathbf{m}
=
\pm
\sqrt{\frac{Q_{N_M}(68.3\%)}{N_M}}
\sqrt{\text{diag} (\tilde{\mathbf{C}}_M)}
,
\label{eq:incert_a}
\end{eqnarray}
defining a confidence interval.
So, firstly, using equation \ref{eq:incert_a} would be better than
using $\pm\sqrt{\text{diag} (\tilde{\mathbf{C}}_M)}$ from a confidence level point of view
(indeed, $\pm\sqrt{\text{diag} (\tilde{\mathbf{C}}_M)}$ is not associated with a constant confidence level $P=68.3\%$ as $N_M$ varies).
But, secondly, equation \ref{eq:incert_a} is still too restrictive, since:
\begin{itemize}
\item
Even in the diagonal case, the full solution of equation \ref{eq:iso_pdf} with $P=68.3\%$ defines a larger (hyper-ellipsoidal) confidence region, 
encompassing the confidence interval defined by equation \ref{eq:incert_a}.
So, equation \ref{eq:incert_a} underestimates uncertainties.
%
%
%
\item
In the general case, $\text{diag} (\tilde{\mathbf{C}}_M)$
is not sufficient to generate a set of admissible tomography models.
Indeed, non-diagonal elements of $\tilde{\mathbf{C}}_M$ can have a strong contribution
as they describe correlations between model space nodes,
which are crucial in tomography
(because of the smoothness, structural conformity, etc. constraints on the model) \cite[]{Duf06}.
Appendix B gives more formal details.
\end{itemize}
So, equation \ref{eq:sol_hyp_elips} must be resolved to represent 
the full solution of equation \ref{eq:iso_pdf}.
It remains to define the $\mathbf{B}$ matrix and how to compute error bars
from the equi-probable contour sampling.

\subsection*{Separating the ``resolved'' space uncertainties}
\label{sec:Eff_null_sec}

We consider an EVD of $\tilde{\mathbf{C}}_M^{-1}$ in the preconditioned domain
(using notations of section ``SVD, EVD...")
\begin{eqnarray}
\mathbf{D}^{\dagger}
\tilde{\mathbf{C}}_M^{-1}
\mathbf{D}^{}
&=&
\mathbf{V}_p
\mathbf{\Lambda}_p
\mathbf{V}^\dagger_p
+
\mathbf{V}_{N_M-p}
\mathbf{\Lambda}_{N_M-p}
\mathbf{V}^\dagger_{N_M-p}
\nonumber\\
&
\underset{
\substack{
\mathbf{\Lambda}_{N_M-p}\rightarrow \epsilon \mathbf{I}_{N_M-p}\\
\text{$\epsilon$ being the noise level}
}
}{
\approx
}
&
\mathbf{V}_p \mathbf{\Lambda}_p \mathbf{V}^\dagger_p + \epsilon \mathbf{I}_{N_M}
.
\label{eq:42}
\end{eqnarray}
%
The second line shares similarity with a damping
and is related to the noise-contaminated effective null space of the tomography.
Equation \ref{eq:42} is a partial EVD of $\tilde{\mathbf{C}}_M^{-1}$ in the preconditioned domain
(all model components or parameters have the same units in this domain), stopped after $p$ iterations when the eigenvalues have reached a fixed prior level $\epsilon$
\cite[]{Zhang1995},
and approximate the effect of the  $N_M-p$ non-computed eigenvectors by $\epsilon \mathbf{I}_{N_M}$.
The damping level $\epsilon$ is tuned by the user
during the tomography pass, as discussed in section ``Tomography inverse problem...''
(remind also that the same priors than the ones of the (last) tomography pass must be used for the posterior covariance computation, as discussed in section ``Bayesian formalism...'').
Using the binomial inverse theorem, Appendix A, we obtain
\begin{eqnarray}
\tilde{\mathbf{C}}_M
&\approx&
\mathbf{D}
\Big[
\mathbf{V}_p \mathbf{\Lambda}_p \mathbf{V}^\dagger_p + \epsilon \mathbf{I}_{N_M}
\Big]^{-1}
\mathbf{D}^{\dagger}
\nonumber\\
&=&
\mathbf{D}
\Big[
\mathbf{V}_p \Big( \mathbf{\Lambda}_p + \epsilon \mathbf{I}_{p} \Big)^{-1} \mathbf{V}^\dagger_p
+ \frac{1}{\epsilon} \mathbf{\Pi}_{null_p}
\Big]
\mathbf{D}^{\dagger}
,
\end{eqnarray}
where $\mathbf{\Pi}_{null_p}$ is the effective null space (dimension $N_M-p$) projector, see section ``SVD, EVD...".
Using equation \ref{eq:sol_hyp_elips},
we can compute
\begin{eqnarray}
\mathbf{B}
&=&
\mathbf{D}
\Big[
\mathbf{V}_p \Big( \mathbf{\Lambda}_p + \epsilon \mathbf{I}_{p} \Big)^{-1/2} \mathbf{V}^\dagger_p
+ \frac{1}{\sqrt{\epsilon}} \mathbf{\Pi}_{null_p}
\Big]
,
\end{eqnarray}
which can be split into
%
\begin{eqnarray}
\label{eq:Bnullcov100}
&&
\mathbf{B}
=
\mathbf{B}^{resolved}_{p} \mathbf{V}^\dagger_p
+
\mathbf{B}^{un-resolved}_{p,N_M}
\\
&&
\mathbf{B}^{resolved}_{p}
=
\mathbf{D}
\mathbf{V}_p \Big( \mathbf{\Lambda}_p + \epsilon \mathbf{I}_{p} \Big)^{-1/2}
\nonumber\\
&&
\mathbf{B}^{un-resolved}_{p,N_M}
=
\frac{1}{\sqrt{\epsilon}} \mathbf{D} \mathbf{\Pi}_{null_p}
\nonumber\\
&&
\quad\quad
\Delta\mathbf{m}=\mathbf{B}\delta\mathbf{r}
\quad\text{where}\quad
\delta\mathbf{r}
\text{ is drawn from a uniform distribution}
\nonumber\\
&&
\hspace{7cm}
\text{and rescaled to have norm $\sqrt{Q_{N_M}(68.3\%)}$}
.
\nonumber
\end{eqnarray}

\begin{figure}[H]
\centering
\includegraphics[width=1.1\linewidth]{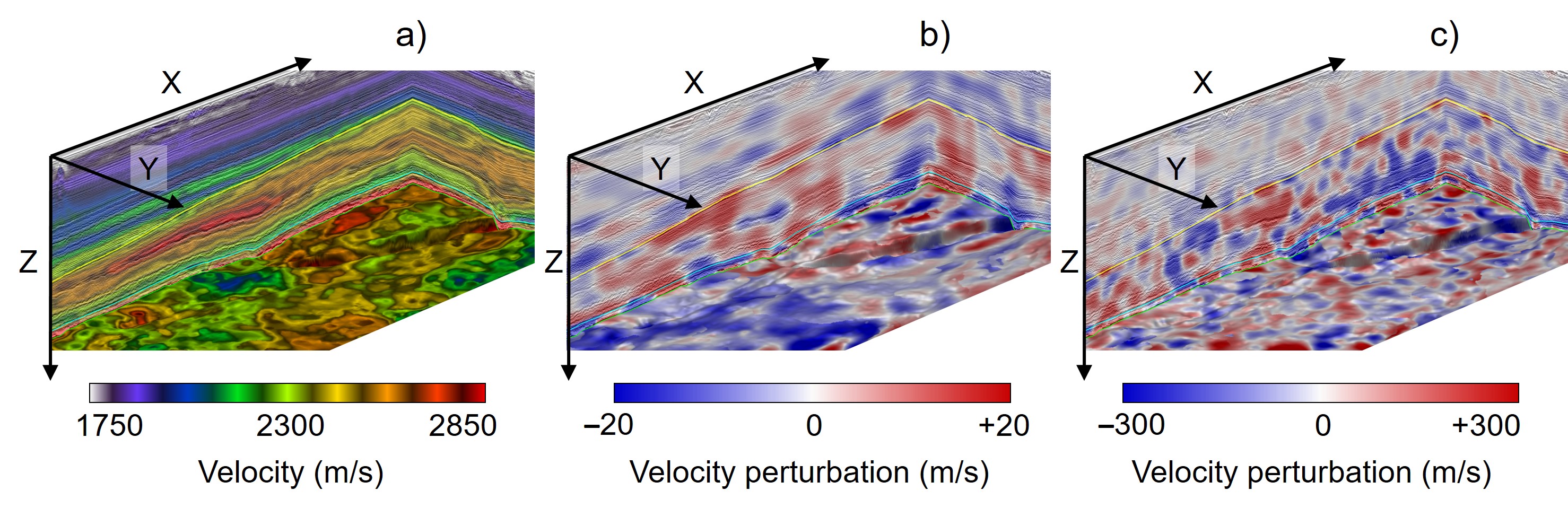}
\caption{
North Sea field data.
Maximum-likelihood velocity (Vp) model $\mathbf{m}_n$.
(a) One resolved space perturbation $\mathbf{\Delta m}^{resolved}$ and (b) corresponding total space perturbation $\mathbf{\Delta m}$, displayed on sections and one horizon.
Modified from \cite{Messud2018_EAGE}; see the article for more details.
}
\label{fig:figure7}
\end{figure}

The method allows us to separate the following two uncertainty contributions:
\begin{itemize}
\item
$\mathbf{B}^{resolved}_{p}$ deals with the effective uncertainty information contained in the posterior covariance matrix.
It drives the contribution to $\mathbf{\Delta m}$ of the eigenvectors with eigenvalues above the prior damping level $\epsilon$, which spans the so-called tomography ``resolved" space (of dimension $p$). As those eigenvectors are greatly constrained by the tomography input data, so are the related resolved space perturbations that are defined by
\begin{eqnarray}
\mathbf{\Delta m}^{resolved}=\mathbf{B}^{resolved}_{p}\mathbf{\delta p}
\quad\text{where}\quad
\mathbf{\delta p}
=
\mathbf{V}^\dagger_p\mathbf{\delta r}
\text{ is a vector of size $p$}
.
\label{eq:proj_p}
\end{eqnarray}
\item
$\mathbf{B}^{un-resolved}_{p,N_M}$ is related to the tomography full effective null space projector
(note that an explicit orthonormalization of the eigenvectors like a Gram-Schmidt can be numerically important 
for an accurate computation of $\mathbf{\Pi}_{null_p}$ \cite[]{Zha07}).
$\mathbf{B}^{un-resolved}_{p,N_M}$ describes the contribution of eigenvectors with eigenvalues below $\epsilon$, which spans the tomography ``unresolved'' space (of dimension $N_M-p$). This space is mostly constrained by the priors and the illumination ($\mathbf{\Pi}_{null_p}$ and $\mathbf{D}$, equation \ref{eq:Bnullcov100}). Unresolved space perturbations are defined by
\begin{eqnarray}
\mathbf{\Delta m}^{un-resolved}=\mathbf{B}^{un-resolved}_{p,N_M}\mathbf{\delta r}
.
\end{eqnarray}
%
%
\item
``Total" perturbations are defined by the sum of both contributions
\begin{eqnarray}
\mathbf{\Delta m}=\mathbf{\Delta m}^{resolved}+\mathbf{\Delta m}^{un-resolved}
.
\end{eqnarray}
\end{itemize}
Despite the formal similarities between the decomposition in equation \ref{eq:Bnullcov100} and the decomposition 
in equation \ref{eq:Bnullcov}, the content is somewhat different:
\begin{itemize}
\item
In equation \ref{eq:Bnullcov}, tomography constraints
are contained in the model preconditioner $\mathbf{P}$ (a steering filter) through equation \ref{eq:dec_p},
whereas in the case of equation \ref{eq:Bnullcov100} the tomography constraints are contained in the eigenvectors
and illumination information is contained in the preconditioner $\mathbf{D}$.
$\mathbf{B}$ is a $N_M\times p$ matrix and
$\mathbf{\delta p}$ a $p$ size vector  in equation \ref{eq:Bnullcov}, whereas
$\mathbf{B}$ is a $N_M\times N_M$ matrix and
$\mathbf{\delta r}$ a $N_M$ size vector in equation \ref{eq:Bnullcov100}.
\item
The $\mathbf{B}^{un-res}_{q,p}$ of equation \ref{eq:Bnullcov}
describes the effective null space
(dimension $p-q$) of the EVD in the model preconditioned domain,
not to be confused with
the $\mathbf{B}^{un-resolved}_{p,N_M}$ of equation \ref{eq:Bnullcov100}
that describes the tomography full effective null space (dimension $N_M-p$).
Starting from equation \ref{eq:Bnullcov100} and considering that the corresponding $p$
is approximately the same as in equation \ref{eq:Bnullcov},
we can deduce a link between the two decompositions at the resolved space level:
\begin{eqnarray}
\mathbf{\Delta m}^{resolved}
=
\mathbf{B}^{resolved}_p\mathbf{\delta p}
&\approx&
(\mathbf{B}^{res}_q+\mathbf{B}^{un-res}_{q,p})\mathbf{\delta p}
=
\mathbf{\Delta m}^{modprec}
\nonumber\\
\mathbf{\delta p}
&=&
\mathbf{V}^\dagger_p\mathbf{\delta r}
.
\label{eq:link}
\end{eqnarray}
This demonstrates that the resolved space perturbations $\mathbf{\Delta m}^{resolved}$, computed with our sheme that works in the full model space, can be related to the perturbations $\mathbf{\Delta m}^{modprec}$ computed with the previously described sheme that works in a model preconditioned domain.
However, important to remark, $\mathbf{\delta p}$ is constrained to be equal to $\mathbf{V}^\dagger_p\mathbf{\delta r}$ in equation \ref{eq:link}.
The introduction of this projection constraint on $\mathbf{\delta p}$
distinguishes our method, also from the one of section ``Cholesky decomposition...''. 
An advantage of considering random perturbations $\mathbf{\delta r}$ of dimension $N_M$
and projecting them on $\mathbf{V}^\dagger_p$ to compute the $p$-dimensional perturbations
$\mathbf{\delta p}$ is that
the latter will already account for information present in the eigenvectors
(structures, etc.),
which will contribute to the resolved space results.\\
Figure \ref{fig:figure7} (b) gives an example of a resolved space perturbation $\mathbf{\Delta m}^{resolved}$ for a field data.
It tends to be organized, smooth (because tomography resolves the large wavelengths of the velocity model), and correlated to structures and to the tomography final model, as expected.
It represents a basis of our uncertainty analysis method.
Remind however that these perturbations represent ``maximum'' perturbations
associated with a $P=68.3\%$ confidence level (samples of the equi-probable contour that bounds the corresponding confidence region), hence the lateral variations visible in figure \ref{fig:figure7} (b).
They should thus not be interpreted like samples of the corresponding normal distribution
(more localized around the maximum likelihood for most of them).
\item
The introduction of $\mathbf{\Delta m}^{un-resolved}=\mathbf{B}^{un-resolved}_{p,N_M}\mathbf{\delta r}$ represents a byproduct specific to the method presented here,
whose computation is little costly once the resolved space computed (i.e. the EVD performed).\\
Figure \ref{fig:figure7} (c) gives an example of a total space perturbation $\mathbf{\Delta m}$ for a field data.
It looks more random and of higher frequency than $\mathbf{\Delta m}^{resolved}$, and has a quite larger magnitude.
Indeed,
$\mathbf{\Delta m}$ is essentially driven by $\mathbf{\Delta m}^{un-resolved}$ as the dimension $N_M-p$ of the unresolved space is quite larger than the dimension
$p$ of the resolved space in most applications.
This, together with the thresholding to $\epsilon$ in equation (\ref{eq:42}),
makes the total space information fully qualitative.
However, it is not without interest. 
It gives an additional information that reflects 
the priors and the illumination,
strongly highlighting the most uncertain gross areas as illustrated further.
It thus can be used for QCs and may also offer some possibility of exploring small-scale non-structural variations, that we cannot consider as fully improbable. 
\end{itemize}

Finally, for completeness, we mention a last element regarding our scheme.
Using notations of section ``Prior-based low-rank ..." we have
\begin{eqnarray}
\mathbf{\delta r}=
\mathbf{R}_{p}\mathbf{\delta r}
+
\mathbf{\Pi}_{null_p}\mathbf{\delta r}
.
\end{eqnarray}
As $\mathbf{\delta p}=\mathbf{V}^\dagger_p\mathbf{\delta r}$ in our scheme,
we can deduce using equation \ref{eq:sol_hyp_elips}
\begin{eqnarray}
\mathbf{\delta r}^\dagger\mathbf{\delta r}
=
\mathbf{\delta p}^\dagger\mathbf{\delta p}
+
\mathbf{\delta r}^\dagger\mathbf{\Pi}_{null_p}\mathbf{\delta r}
=
Q_{N_M}(68.3\%)
.
\end{eqnarray}
In the large $N_M$ case (like in tomography),
where $Q_{N_M}(68.3\%)$ remains close to $N_M$,
one may think the latter equation implies
$\mathbf{\delta p}^\dagger\mathbf{\delta p}= p$
and
$\mathbf{\delta r}^\dagger\mathbf{\Pi}_{null_p}\mathbf{\delta r}= N_M - p$,
which would help to slightly simplify the sampling problem in equation \ref{eq:Bnullcov100}.
However, this is not the case as the
amplitudes of the perturbations in resolved and unresolved subspaces
are not fully independent from the total uncertainty point of view.

\subsection*{Computing error bars}
\label{sec:stat}

\begin{figure}[H]
\centering
\includegraphics[width=0.8\linewidth]{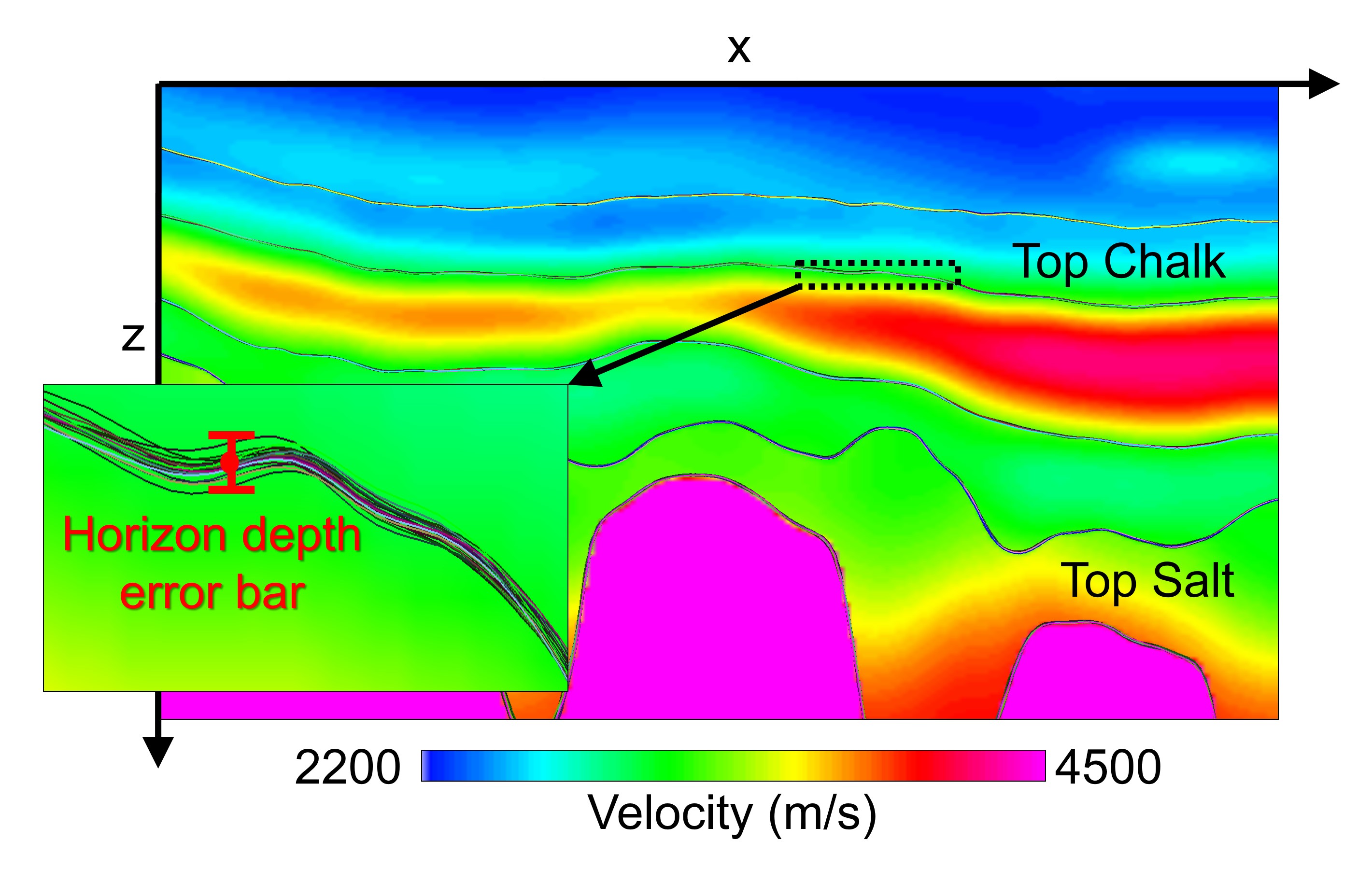}
\caption{
North Sea field data.
The tomography maximum-likelihood velocity model overlain with a subset of 20 (among several hundred) horizons perturbations, obtained by map migrations in perturbed velocity models (resolved space).
Modified from \cite{Messud2017_TLE}; see the article for more details.
}
\label{fig:figure2}
\end{figure}

Uncertainty attributes can be computed statistically using the obtained set of perturbations, especially for the resolved space using the $\{\mathbf{\Delta m}^{resolved}\}$, and also possibly for the total space using the $\{\mathbf{\Delta m}\}$:
\begin{itemize}
\item
Tomography model $68.3\%$ error bars on $\mathbf{m}_n$ (velocity and anisotropy models):\\
Computed by considering the maximum possible variations of the model perturbations ($i$ here represents the model grid coordinates):
\begin{eqnarray}
\sigma^{(\mathbf{m_n})}_i=\max\{\Delta m_i\}
.
\end{eqnarray}
The true model belongs to $\mathbf{m_n}\pm\mathbf{\sigma}^{(\mathbf{m_n})}$ with a probability $P \ge 68.3\%$ \cite[]{Messud2017_EAGE}.
\cite[]{Reinier2017_EAGE} give an illustration of such error bars for the velocity and the total space.
\item
Maximum horizon perturbations within the $68.3\%$ confidence region:\\
Map migrations can be performed using each model
perturbation to obtain a set of equi-probable horizon perturbations $\{\mathbf{\Delta h}\}$ 
around the maximum-likelihood horizon position $\mathbf{h_n}$,
remind section ``Migration structural uncertainties''.
Note that those perturbations must not be interpreted as a migration structural posterior pdf sampling,
but as a sampling of the pdf's $68.3\%$ equi-probable contour, 
producing maximum possible 
perturbations within the $68.3\%$ confidence region.
The perturbations can be QCed, see Figure~\ref{fig:figure2} for a field data example,
and used to compute horizon error bars (next item),
but they should not be used as such in reservoir workflows that need
perturbed horizons sampled from the pdf as an input
(as the latter samples are not similar, tending to be
more concentrated around the maximum-likelihood).
If needed, pdf sampling can easily be recovered from equi-probable perturbations
(a Gaussian pdf can easily be reconstructed from one of
its equi-probable contours, cf. section ``Sampling a Gaussian equi-probable contour...").
\item
Horizon position $68.3\%$ error bars:\\
Depth error bars can be defined as the maximum possible depth variation
of the horizon perturbations
($i$ here represents the horizon coordinates):
\begin{eqnarray}
\sigma^{(\mathbf{h_n})}_i=\max\{\Delta h_i\}
.
\end{eqnarray}
The horizon migrated points ``move'' vertically and laterally for each map migration;
$\sigma^{(\mathbf{h_n})}_i$ considers the depth ``envelope" of all migrated horizons and thus accounts for lateral displacements of migrated points. 
This gives $68.3\%$ confidence level error-bars:
the true horizon depth position belongs to $\mathbf{h_n}\pm\mathbf{\sigma}^{(\mathbf{h_n})}$ with a probability $P \ge 68.3\%$ \cite[]{Messud2017_EAGE}.
Lateral (x and y-directions) horizon error bars can be computed using the same principle, from differences of position between rays traced in $\mathbf{m}_n$ and rays traced in the perturbed model.
\end{itemize}

Note that the error bars ($\sigma^{(\mathbf{m_n})}$ and $\sigma^{(\mathbf{h_n})}$) should not be computed from standard-deviations of the perturbations but from a maximum,
as we sample a pdf equi-probable contour.
Also, our error bars definitions account for the non-diagonal elements of the tomography and migration structural posterior covariances, equations \ref{eq:post_pdf2bis} and \ref{eq:macro_t_cov}.
Thus, they contain more uncertainty information than the diagonal elements of $\tilde{\mathbf{C}}_M^{1/2}$ and $\tilde{\mathbf{C}}_H^{1/2}$ (i.e. the standard-deviations).
They therefore can be considered as ``generalized standard deviations".

All previous error bars can be computed statistically the same way for the total space,
using the $\{\mathbf{\Delta m}\}$ perturbations.
However, as the latter are fully qualitative, it then seems difficult to keep a $68.3\%$ confidence level interpretation.
However, the hierarchy of the total space are not without interest as illustrated further.

The computational cost of the method resides in an EVD of the final tomography operator
(equations \ref{eq:A-gA-g+} and \ref{eq:42})
and map migrations of horizons in perturbed models
(stable error bars are with 200-500 random models as the equi-probable contour sampling
reduces the sampled space to its most representative components).
It approximately equals the computational cost of one additional non-linear slope tomography iteration.

\subsection*{Specific to non-linear slope tomography: Migration volumetric positioning error bars and QC of the Gaussian hypothesis}
\label{sec:QCgauss}

The use of non-linear slope tomography, based invariants picks, provides unique advantages:
\begin{itemize}
\item
Non-linear slope tomography provides an efficient way to assess the quality of the randomly generated model perturbations. 
Indeed, the cost functions related to the perturbations can be estimated automatically and non-linearly
by kinematic migration of the invariant picks, for an affordable additional computational cost 
(this step involves computationally effective ray tracing only).
Combined with the posterior pdf equi-probable contour sampling, it allows us 
to QC the validity of the Gaussian hypothesis done in section ``Bayesian formalism...".
Indeed, these cost function values must equal the left side of equation \ref{eq:iso_pdf} (up to an additive constant) in the linear approximation.
In other terms, if the linear and Gaussian approximations are pertinent, computed perturbations should be equi-probable or equivalently iso-cost.
This will be illustrated further.
\item
As discussed in section ``Migration structural uncertainties",
full kinematic migrations (using all offsets) of the invariant picks (using all data, not only target reflectors invariant picks)
can be performed on each model perturbation.
We compute how much a migrated pick has moved compared to its position when kinematically migrated in the maximum likelihood model (both being related to the same invariant pick).
Then, using a ``maximum" principle as in section ``Computing error bars...''
allows us to deduce migration error bars at all migrated picks positions, not only at horizons positions.
Interpolating corresponding error bars give error bars in the whole migrated volume,
called migration volumetric $68.3\%$ error bars in the following.
This makes it possible to deduce positioning uncertainty quantities
in the whole migrated volume, not only at target reflector positions.
(We remind the error bars should be interpreted
in the light of the data that have been used to compute them,
here reflection events related to the migrated picks.)
These error bars will be illustrated further.
\end{itemize}

\subsection*{Can computed uncertainties be more than qualitative?}
\label{sec:glob}

We already mentioned that the results discussed in this paper depend on the
approximations of the covariance matrices $\mathbf{C}_D$
(quality of the invariant picks)
and $\mathbf{C}_M$
(various model space constraints) that enter into $\tilde{\mathbf{C}}_M$ computation,
leading to somewhat qualitative uncertainty evaluation in practical applications.
Also, it is not possible to account for all possible sources of uncertainties,
but only for some of them that we consider as the most important (for instance here, the tomography model is considered as a main source of uncertainty for the migrated domain).
This reinforces the somewhat qualitative feature of the uncertainty evaluation.

\begin{figure}[H]
\centering
\includegraphics[width=0.4\linewidth]{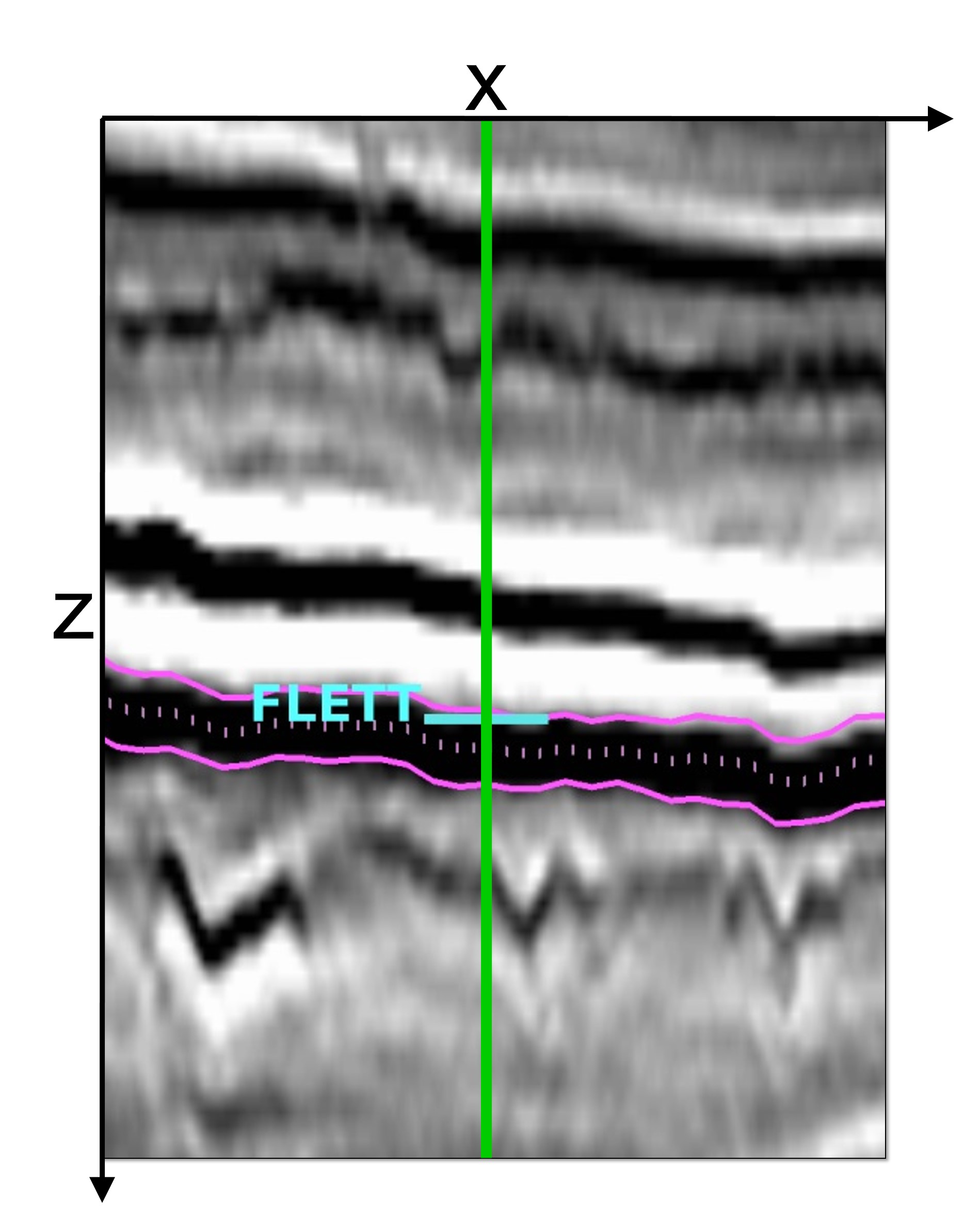}
\caption{
Zoom-in around one well with Base Flett marker (cyan), migrated image section, Base Flett horizon (pink dotted line) and horizon depth error bars schematized by the pink lines.
Modified from \cite[]{Reinier2017_EAGE}.
}
\label{fig:figure6}
\end{figure}

Additionally, many aspects that only slightly affect the maximum-likelihood search
(i.e. the minimum of the cost function)
may affect the uncertainty computation
(i.e. the curvatures of the cost function at the minimum).
Often overlooked,
those aspects tend to  affect uncertainties up to a global proportionality constant:
\begin{itemize}
\item
Bayesian uncertainty reasoning holds strictly if the diagonals of $\mathbf{C}_D$
and $\mathbf{C}_M$, that enter into $\tilde{\mathbf{C}}_M$ computation,
represent variances, i.e. are related to a confidence level of $68.3\%$.
In practice, even good approximations of $\mathbf{C}_D$ and $\mathbf{C}_M$
tend to be defined up to a global proportionality constant,
i.e. they are balanced together, so that they do not affect the maximum-likelihood.
But the scaling of the global proportionality constant
is not easy and itself uncertain.
\item
Data decimation will produce less ``illumination" of each model node and therefore will tend to
increase the uncertainties.
Contrariwise, a larger model discretization step will produce more ``illumination" of each model node and thus will tend to
decrease the uncertainties.
Those effects could theoretically be compensated by fine adaptation of the prior covariances,
but this is not easy and basically requires knowledge of a large part of the inversion solution.
Reasonable changes in data decimation and model discretization will tend to affect the uncertainties globally and linearly on average, i.e. up to a global proportionality constant.
\end{itemize}
The combination of these two effects will tend to  affect uncertainties approximately only up to a global proportionality constant.
%
This proportionality
constant can be rescaled
using posterior information external to the tomography, like wells,
%
so that all well markers lie within the horizon error bars,
see illustration in Figure \ref{fig:figure6}.
The resolved space error bars should of course be used for such a matching, as they contain
the tomographic operator information; after rescaling they become less qualitative or more quantitative.
However, they still will remain somewhat qualitative because of the
various previously mentioned  caveats.
This is true for all methods discussed in this article.

The total space error bars always remain fully qualitative,
among others as the null space exploration has been thresholded at the tomography noise level.
They may be rescaled
by the global constant found from the resolved space error bars (as is the case in next illustrations),
but only their hierarchy is to be considered (not their values) as a complementary information.

\section*{Illustration on synthetic data}
\label{sec:resultssynth}

\begin{figure}[H]
\centering
\includegraphics[width=0.9\linewidth]{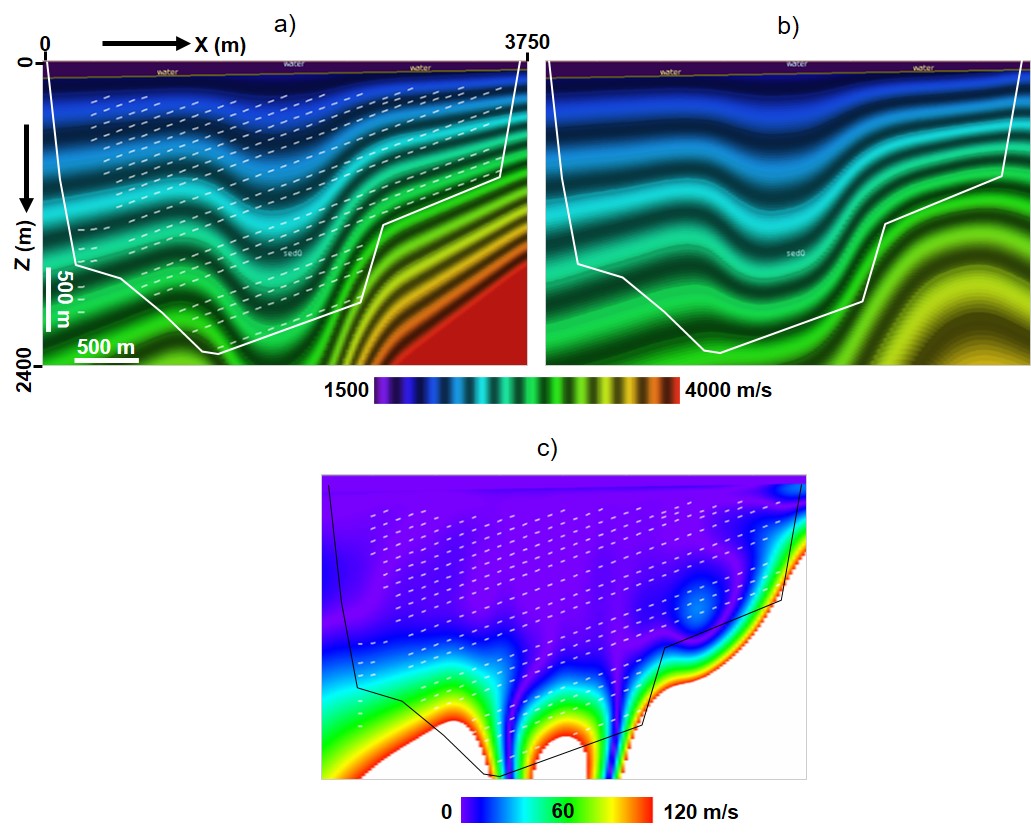}
\caption{
Synthetic 2D velocity model. White small line segments represent locally coherent reflector events (related to migrated picks) and white/black line delineates the area illuminated by the tomography data. (a) True model $\mathbf{m}^*$. (b) Best model $\mathbf{m}_n$ obtained after 20 internal nonlinear iterations of slope tomography having started with constant velocity model ($\mathbf{m}_0=2300$ m/s).
(c)
Absolute difference $|\mathbf{m}_n-\mathbf{m}^*|$ (the absolute value is taken element-wise) between true and best inverted velocity models.
}
\label{fig:figuresynt1}
\end{figure}

We consider a synthetic 2D case to illustrate the previously described method.
Noise has been added to the exact invariant picks for more realism.
We consider perturbations of the velocity model (Vp)
(but, as previously mentioned, the formalism is general for any subsurface model including anisotropy).
Figure \ref{fig:figuresynt1} (a) shows the true velocity model $\mathbf{m}^*$ with superimposed segments that represent locally coherent reflector events related to migrated picks, remind figure \ref{fig:figure_tomo}.
Figure \ref{fig:figuresynt1} (b) shows the best velocity model $\mathbf{m}_n$ obtained after 20 iterations of non-linear slope tomography, having started with a constant velocity model.
Inside the area delineated by the black/white line in figure \ref{fig:figuresynt1},
that schematizes the area illuminated by rays connected to the invariant picks, the tomography result is good but with some differences with the exact model.
These are highlighted in figure \ref{fig:figuresynt1} (c), representing $|\mathbf{m}_n-\mathbf{m}^*|$ (where the absolute value is taken element-wise).
Beyond the black/white line, the tomography is not expected to recover the model.

\begin{figure}[H]
\centering
\includegraphics[width=0.9\linewidth]{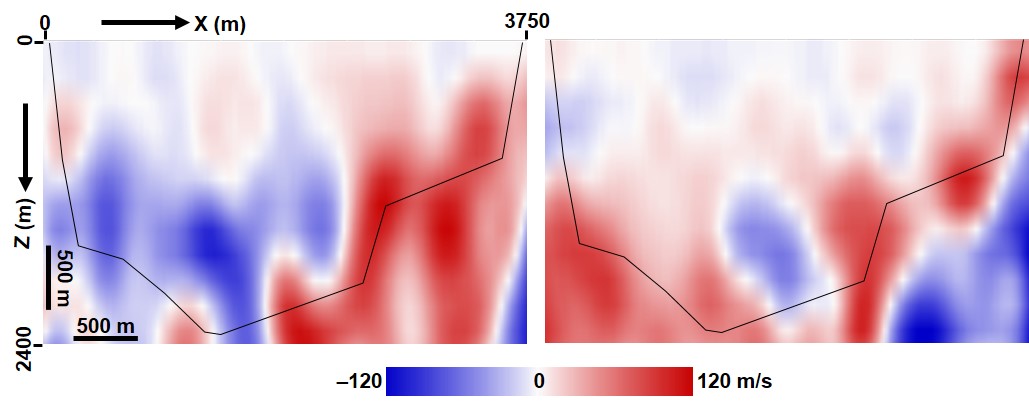}
\caption{
Two resolved space equi-probable velocity perturbations (i.e. computed along the same posterior pdf contour) on $\mathbf{m}_n$.
Black line delineates the area illuminated by the tomography data.
}
\label{fig:figuresynt6}
\end{figure}

\begin{figure}[H]
\centering
\includegraphics[width=0.9\linewidth]{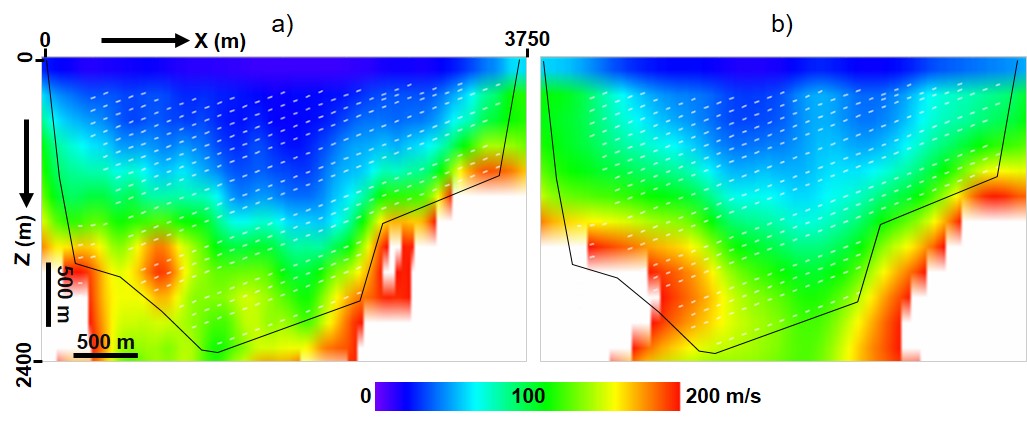}
\caption{
Different types of velocity error bars on $\mathbf{m}_n$. (a) Resolved space velocity error bars. (b) Velocity standard-deviation estimated by the LSQR solver of tomographic matrix (rescaled). Black line delineates the area illuminated by the tomography data.
}
\label{fig:figuresynt3}
\end{figure}

We now apply the previously described uncertainty analysis method, where
$\mathbf{m}_n$ is considered as the maximum likelihood model.
Figure \ref{fig:figuresynt6} shows two resolved space equi-probable velocity perturbations
and figure \ref{fig:figuresynt3} (a) the resolved space velocity error bars.
Comparing these results magnitudes to figure \ref{fig:figuresynt1} (c),
we observe that the true velocity model $\mathbf{m}^*$ lies within the range of perturbations of and error bars on $\mathbf{m}_n$, which is satisfying.

Also, figure \ref{fig:figuresynt3} (b) shows the (rescaled) standard-deviation estimated by the LSQR solver of tomography. It exibits in this simple case relatively similar variations than the ones of the resolved space velocity error bars, figure \ref{fig:figuresynt3} (a).
This is satisfying as, even if our error bars contain more uncertainty information than only standard-deviations (related to diagonal elements of the posterior covariance), the standard-deviations tend to represent a first order contribution in many applications, especially in simple synthetic cases where non-local constraints like a structural ones were are not activated.
However, note that the resolved space error bars contains a more subtle hierarchy,
related to the correlations in $\tilde{\mathbf{C}}_M$,
depending on the contributions of the noise in the invariant picks and
the form of the tomography modeling Jacobian matrix.
We call this the uncertainty related to the ``tomography discrimination power''.

\begin{figure}[H]
\centering
\includegraphics[width=1.\linewidth]{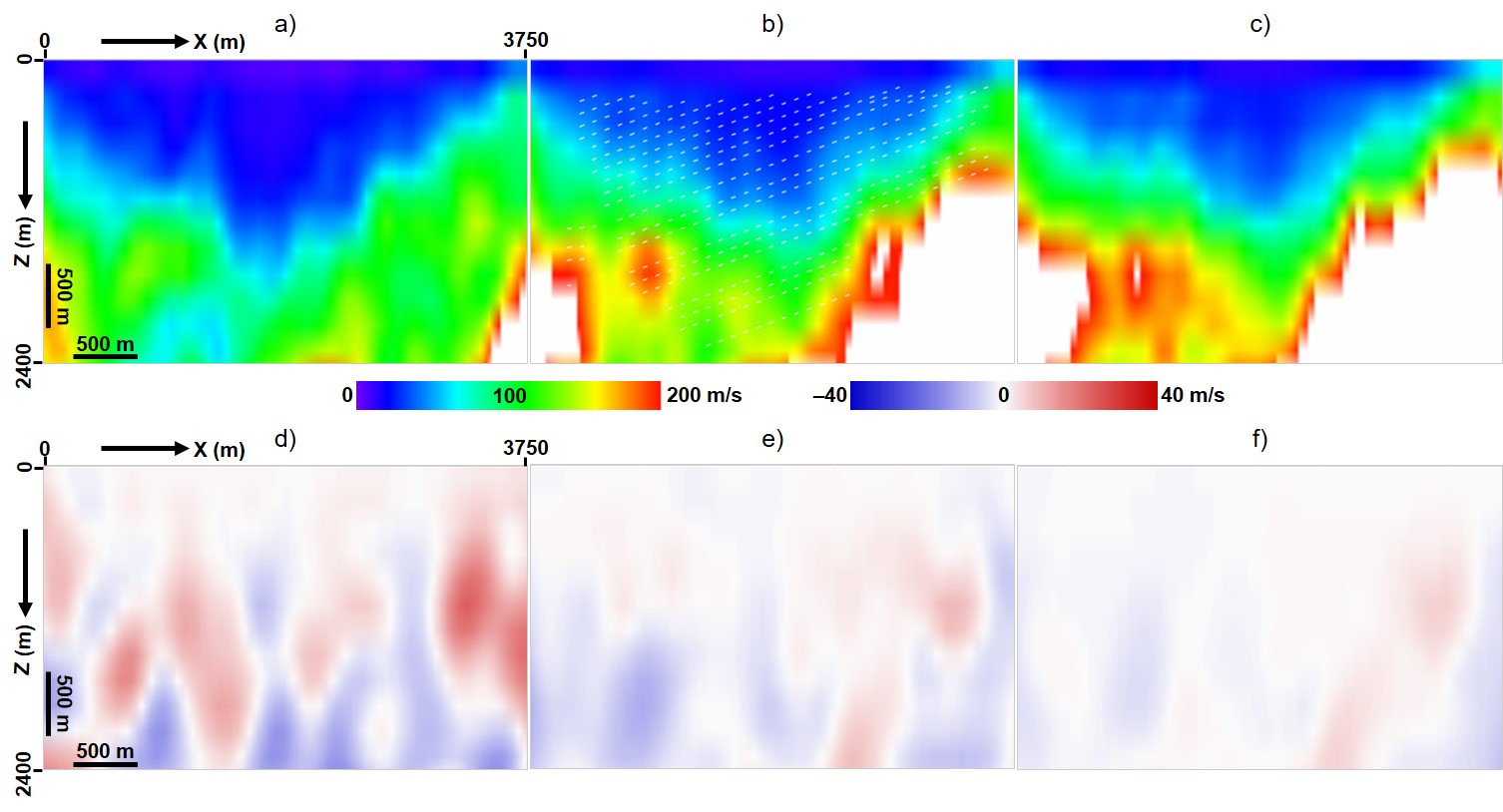}
\caption{
Variations of mean velocity perturbations and of velocity error bars with the number of perturbations (resolved space). The velocity error bars are represented for (a) 20 perturbations, (b) 200 perturbations and (c) 900 perturbations. Similarly, the mean velocity perturbations are represented for (d) 20 perturbations, (e) 200 perturbations and (f) 900 perturbations. 
}
\label{fig:figuresynt5}
\end{figure}

Figure \ref{fig:figuresynt5} shows the variations of mean velocity perturbations and of velocity error bars with the number of velocity perturbations, for the resolved space. 
The mean of the velocity perturbations tend towards zero when the number of realizations increases, as expected.
The velocity error bars stabilize when the number of samples increases and we observe that 200 samples represents a good compromise.
This is a typical value, also for large scale applications.

\begin{figure}[H]
\centering
\includegraphics[width=1.\linewidth]{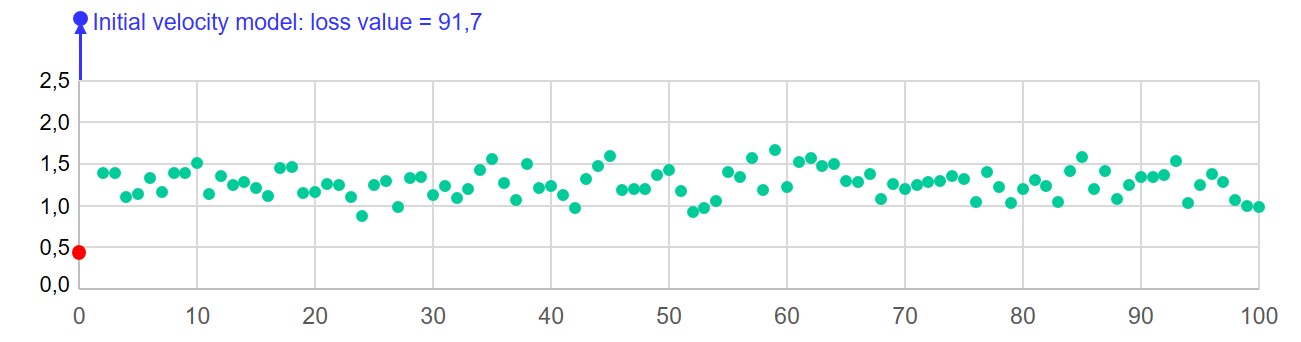}
\caption{
Cost function values (in green) obtained by kinematic migration of invariant picks in the 
first 100 velocity perturbations (resolved space).
Mean cost function value of all perturbations is 1,27. 
Cost function value for the tomography  model is 0,48 (red dot)
and 91,7 for initial model (constant velocity of 2300 m/s, blue dot).
}
\label{fig:figuresynt7}
\end{figure}

Figure \ref{fig:figuresynt7} represents the non-linearly computed tomography cost function values 
(by kinematic migration of invariant picks)
in 100 resolved space velocity perturbations.
As discussed in section ``Specific to non-linear slope tomography...'', if the linear and Gaussian approximations assumed in the uncertainty analysis are appropriate, perturbations should be iso-cost.
We observe in figure \ref{fig:figuresynt7} that the perturbations tend to be iso-cost, with only limited dispersion around the average of the cost function values of the perturbed models.
This QC, specific to our method, allows to validate the linear and Gaussian hypothesis.
Interestingly, we also see that no spurious perturbations related to too large variations of the cost function were generated;
it is thus not necessary to have an additional step discarding spurious perturbations 
with the method discussed here (whereas it may be necessary with the method of section ``Sampling a normal distribution...").

Note that the dispersion range of the cost function values in the velocity perturbations depends on the shape of the cost function valley around the best model,
decorrelated from the best model cost function value. 
In synthetic cases, where the problem is well constrained, 
the best model cost function is very low and
the global minimum valley is sharp.
This leads to more ``uniqueness'' of the solution and favors non-linearity, thus tends usually to produce a larger dispersion range for the cost function values in the perturbations.

\begin{figure}[H]
\centering
\includegraphics[width=0.9\linewidth]{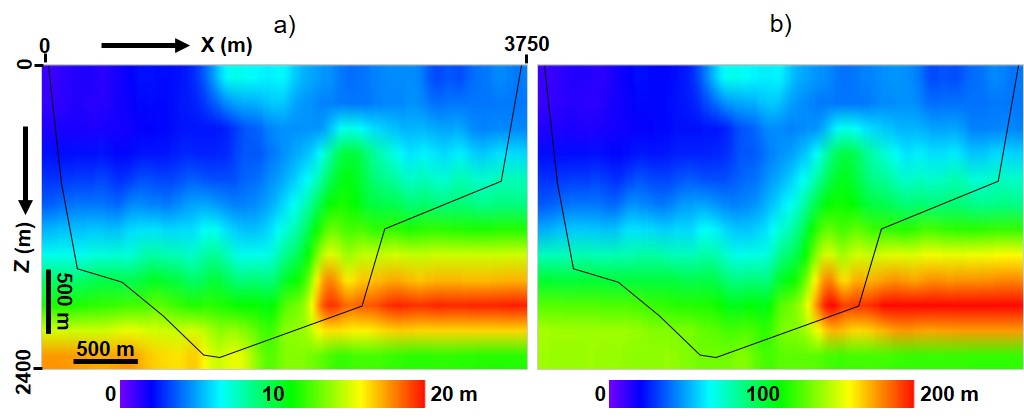}
\caption{
Migration ``volumetric'' depth error bars (resolved space). Effect of a proportionality constant equal to $1$ (a) and $10$ (b) applied in the velocity domain.
Black line delineates the area illuminated by the tomography data;
outside this area, the error bars are just extrapolated until the edges,
so that they are meaningless.
The scaling factor in the velocity domain translates into the same scaling factor in the migrated domain, confirming that the linear hypothesis is valid for the range of perturbations.
}
\label{fig:figuresynt4}
\end{figure}

We now interest ourvelves to error bars in the migrated domain.
In this synthetic case, it would not be pertinent to define an horizon and compute its perturbations by map migrations.
This computation will rather be illustrated in next section on field data where target horizons make sense.
So, we propose here to check the migration ``volumetric'' depth error bars, specific to our method, shown in figure \ref{fig:figuresynt4} (a).
These uncertainties make sense only inside the area illuminated by the tomography data, delineated by the black line.
Outside this area, the migration ``volumetric'' depth error bars are just extrapolated until the edges,
so that they are meaningless.
These error bars allow to quantify how the perturbations of the best model, like in figure \ref{fig:figuresynt6}, as well as the ``velocity stall'' in the best model,
figure \ref{fig:figuresynt1} (b), affect the migrated space (here related to the migrated picks depths).

The migration ``volumetric'' depth error bars also allow to verify that the linear hypothesis assumed in the analysis for the migrations is pertinent, complementarily to figure \ref{fig:figuresynt7} that allowed to verify that the linear hypothesis assumed in the velocity domain is pertinent.
Figure \ref{fig:figuresynt4} shows the effect of a proportionality constant equal to $1$ (a) and $10$ (b) applied in the velocity domain.
The scaling factor in the velocity domain almost only translates into the same scaling factor in the migrated domain.
This allow to QC that the linear hypothesis for the migrations is valid for our range of perturbations.

\begin{figure}[H]
\centering
\includegraphics[width=0.9\linewidth]{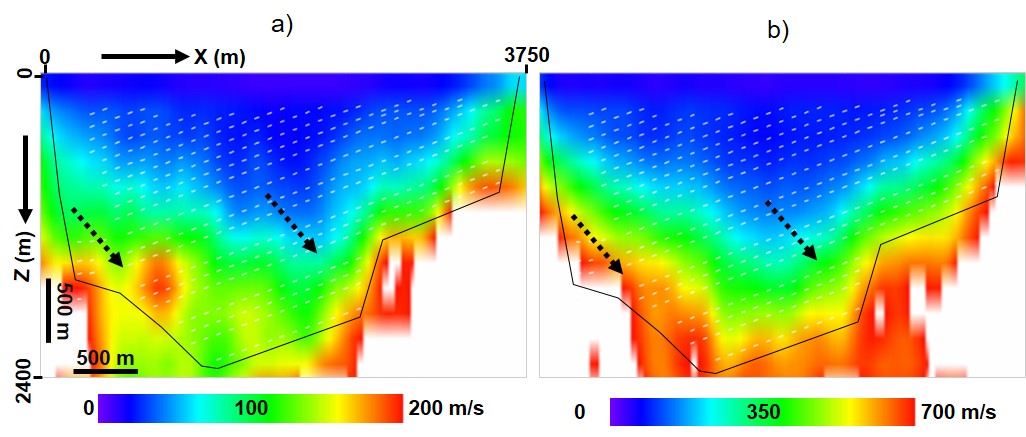}
\caption{
Velocity error bars. (a) Resolved space and (b) total space. 
Black line delineates the area illuminated by the tomography data.
Dashed black arrows highlight some difference in the hierarchy of resolved and total space error bars
in the illuminated area.
}
\label{fig:figuresynt10}
\end{figure}

Finally, we discuss the total space by-product.
Figure \ref{fig:figuresynt10} compares the resolved and total spaces velocity error bars.
The first thing we can note is that the total space error bars are here approximately 
three times larger than the resolved space error bars.
This is because the dimension $N_M-p$ of the unresolved space is here approximately three times larger than the dimension $p$ of the resolved space; this makes the unresolved space perturbations dominate the total space perturbations.
Anyway, the total space error bars are fully qualitative so that only their hierarchy is to be considered, not their values.

Even if this synthetic case is simple and only allows to highlight subtle differences between total and resolved space error bars, we observe in figure \ref{fig:figuresynt10}
that the total space error bars delineate more precisely the area illuminated by the tomography data, i.e. the black line, than the resolved space error bars.
The total space error bars may serve as a complementary QC,
identifying areas where computing error bars is meaningless (especially the much less illuminated areas),
this information being ``attenuated'' in the resolved space error bars
as they concentrate more on the uncertainty related to the tomography discrimination power. 
Doing so, the latter give a more subtle hierarchy in the reasonnably illuminated areas,
highlighted by the dashed black arrows in figure \ref{fig:figuresynt10}.
The resolved space error bars are more quantitative (even if still somewhat qualitative) and
usefull for futher use in decision-makings.

\section*{Error bars illustrated on 3D field data}
\label{sec:results}

Industrial applications of the method presented in this paper have been published in
\cite{Messud2017_EAGE, Messud2017_TLE,Messud2018_EAGE}, \cite{Reinier2017_EAGE} and \cite{Coleou2019_EAGE}.
In this section, we gather few examples illustrating error bars in the migrated domain.
For further details, we invite the reader to refer to the aforementioned articles.

\begin{figure}[H]
\centering
\includegraphics[width=1.\linewidth]{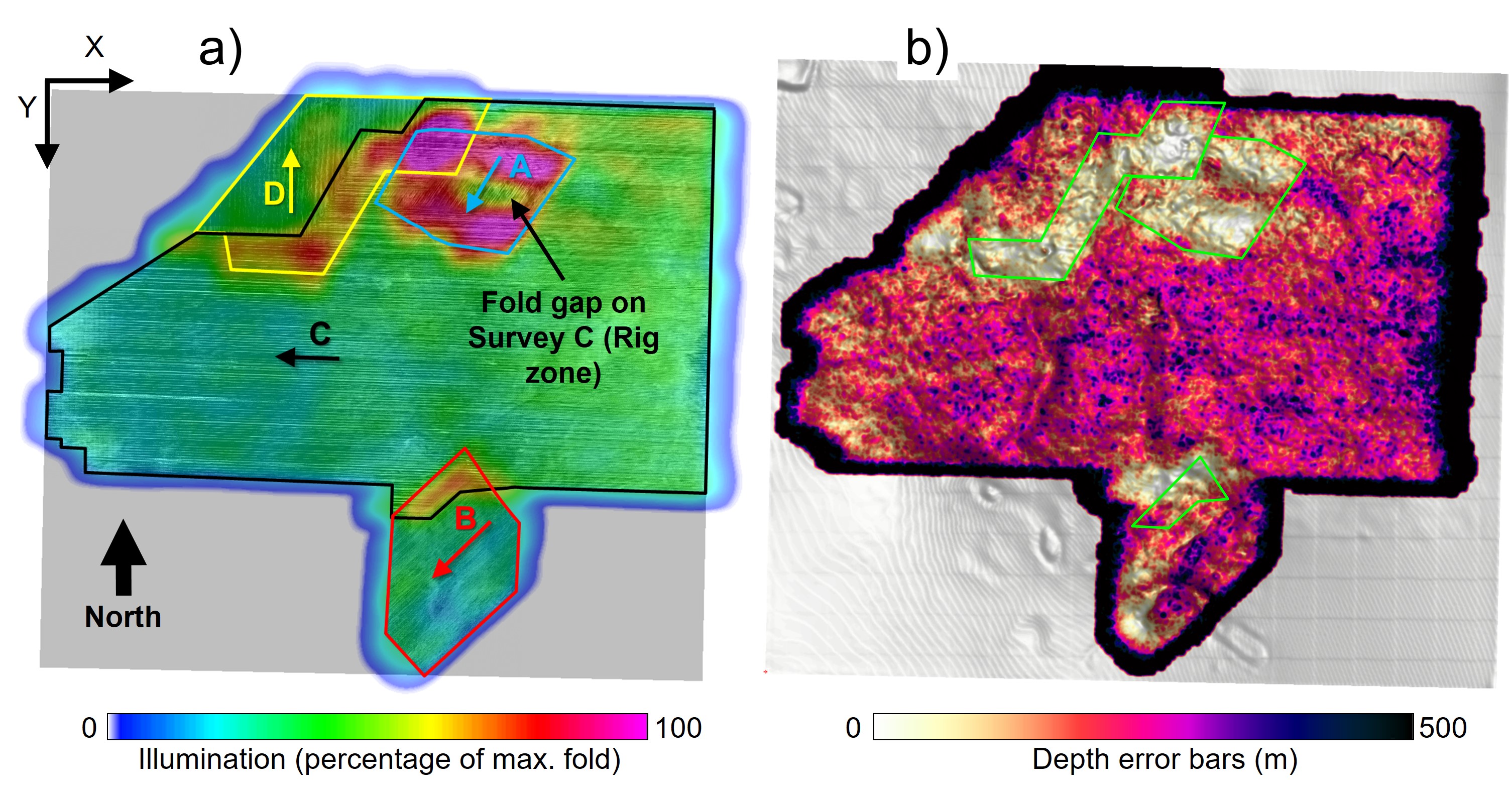}
\caption{
Top Chalk horizon. (a) Illumination map with shooting direction of each survey indicated by the direction of the associated arrow.
(b) Total space horizon depth error bars.
Modified from \cite{Messud2017_EAGE}; see the article for more details.
}
\label{fig:figure3}
\end{figure}

Let us illustrate our method on a first North Sea dataset
merging four different overlapping narrow-azimuth towed-streamer surveys acquired over the years with different layout configurations.
Figure \ref{fig:figure3} (a) shows a compounded fold map of the surveys labelled A to D.
The arrows indicate the different acquisition directions, and the overlapping parts with higher fold appear clearly. 
Figure \ref{fig:figure3} (b) displays the computed total space depth error bars for the top chalk horizon. 
One can observe a clear correlation between the illumination map and the total space depth error bars: 
the latter are smaller in overlap areas where the angle diversity (dip and azimuth) of raypaths is larger. 
On the other hand, lower-fold areas such as the rig zone inside survey C show relatively higher total space  error bars correlated with the poorer angle diversity of raypaths and the reduced illumination.
We can also observe much larger error bars on poorly illuminated survey edges. 
So, total space error bars highlight the combined effects of the acquisition fold and of the effective angle diversity that is in particular sensitive to structural complexities.
These error bars are fully qualitative but give an information related to illumination issues
that can be usefull as a complementary QC.
Indeed, they allow to better identify where computing error bars is meaningless,
here mostly on the edges,
this information being ``attenuated'' in the resolved space error bars as illustrated by the next example.

\begin{figure}[H]
\centering
\includegraphics[width=1.1\linewidth]{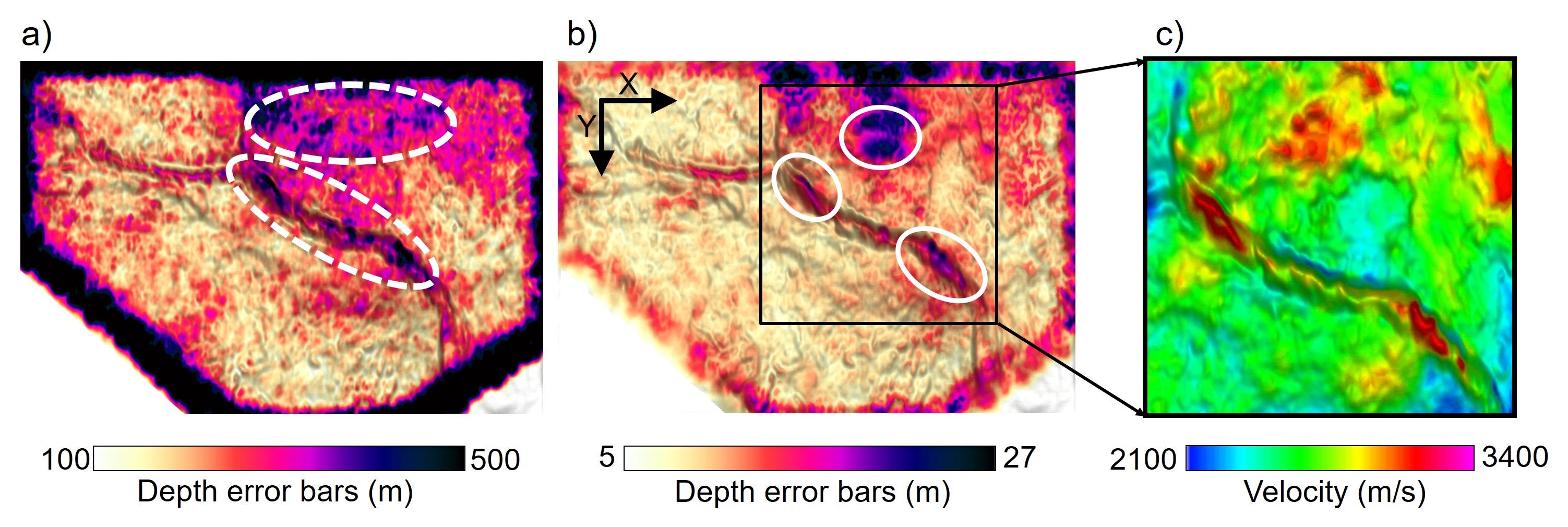}
\caption{
Horizon depth error bars. (a) Total space, (b) resolved space and (c) velocity (Vp), extracted above an horizon.
Modified from \cite{Messud2018_EAGE}; see the article for more details.
}
\label{fig:figure8}
\end{figure}

Figure \ref{fig:figure8} (a) shows total horizon depth error bars.
Again, they highlight the acquisition illumination variations,
allow to clearly identify where computing error bars is meaningless,
again mostly on the edges.
This information is strongly ``attenuated'' in the corresponding resolved space depth error bars,
figure \ref{fig:figure8} (b).
Within the area where computing resolved space error bars make sense, 
the resolved space error bars give a more detailed hierarchy,
correlated to the uncertainty in the tomography discrimination power.
Figure \ref{fig:figure8} (b) highlights the larger resolved space error bars in steeply dipping parts of the top Chalk horizon located below velocity features in the overburden. Comparing with figure \ref{fig:figure8} (c), we can observe the correlation between the velocity features in the overburden and the spatial distribution of the resolved space depth error bars at top chalk level.
Also illustrated in the figure 4 in \cite{Messud2017_TLE},
resolved space error bars correlate very well with steeply dipping flanks or faults, and are stronger when shooting and dipping/fault plane directions are parallel, thus confirming that shooting along the dip direction is better for resolution than shooting strike.
The resolved space error bars are more quantitative and
usefull for futher use in decision-making and risk mitigation.

\begin{figure}[H]
\centering
\includegraphics[width=0.6\linewidth]{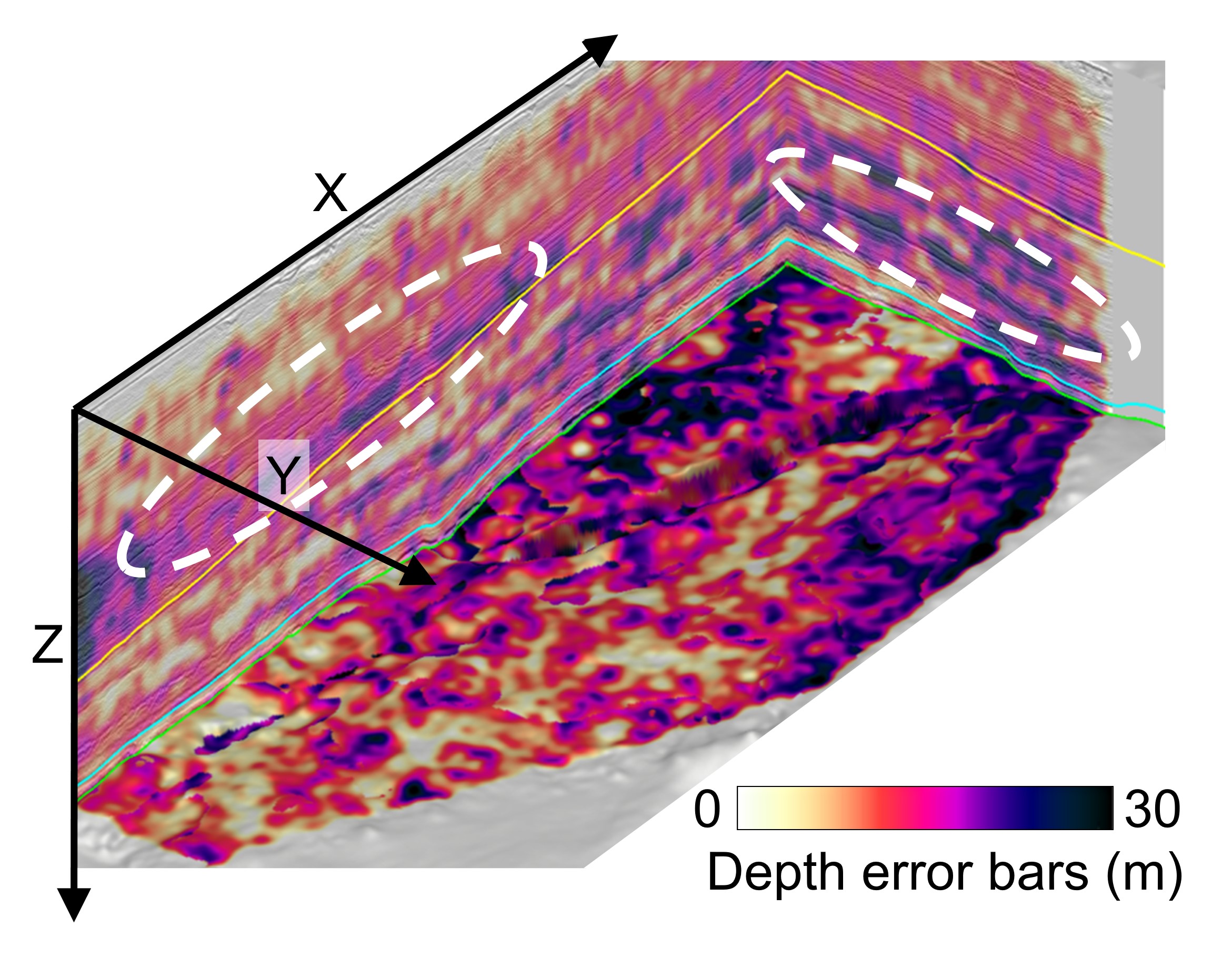}
\caption{
Migration volumetric depth error bars (resolved space), displayed on vertical sections and on a horizon.}
\label{fig:figure10}
\end{figure}

In these examples, horizon error bars were computed by map migration in model perturbations. 
However, as discussed in section ``Specific to non-linear slope tomography...", 
non-linear slope tomography allow computing migration volumetric error bars by full kinematic migration (using all offsets) of the invariant picks in model realizations.
Figure \ref{fig:figure10} shows an example of migration volumetric depth error bars for the resolved space, extracted on vertical sections and along a horizon. 
These error bars exhibit layered and velocity-correlated variations having longer spatial wavelengths that the horizon error bars, as the full-offset range (not only zero-offset) is considered in the kinematic migrations.
The advantage of the volumetric error bars is that they make it possible to track and understand the buildup of positioning uncertainties in the overburden
and in-between horizons. 

\begin{figure}[H]
\centering
\includegraphics[width=0.9\linewidth]{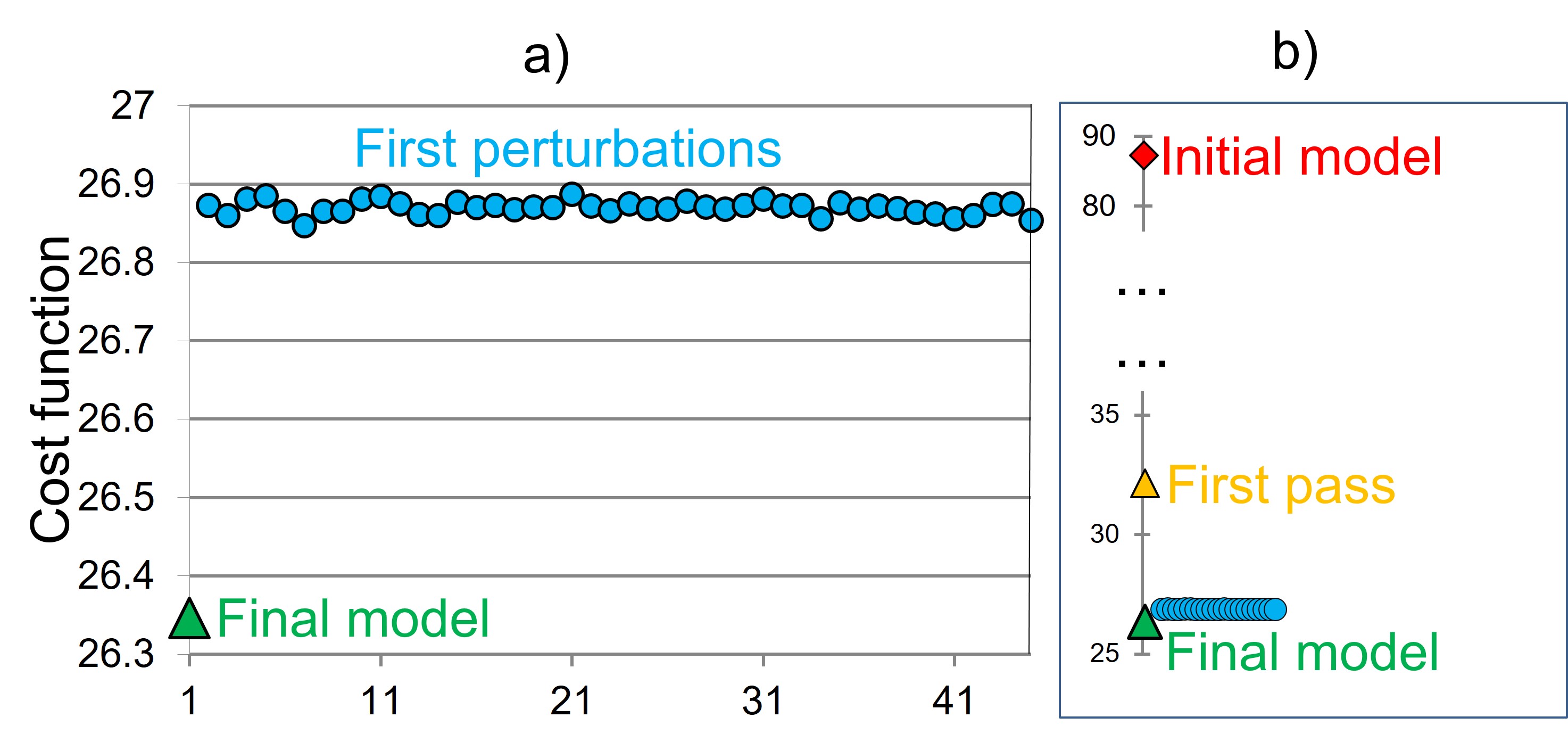}
\caption{
North Sea field data.
(a) Tomography cost function values for the final model (i.e. maximum-likelihood model) and the first 46 generated perturbed models among 500. (b) Cost function values after the first tomography pass and the final tomography pass are represented with a larger nonlinear scale.
Modified from \cite{Reinier2017_EAGE}.
}
\label{fig:figure5}
\end{figure}

Figure \ref{fig:figure5} shows, for first 46 resolved space perturbations,
the non-linearly computed tomography cost functions.
Almost iso-cost, i.e. equi-probable, perturbations were generated. We observe limited variations around the average of the cost function values of the perturbed models, meaning that the Gaussian and linear hypothesis assumed in the analysis is appropriate.
Again, we see that no spurious perturbations related to too large variation of the cost function were generated.
It is not necessary to have an additional step discarding spurious perturbations 
with the method discussed here (whereas it may be necessary with the method of section ``Sampling a normal distribution...").

Let us now illustrate the integration of structural uncertainties into a downstream gross-rock volume (GRV) calculation workflow on a second seismic dataset.
In a conventional stochastic approach, structural uncertainties are known at well locations and inferred elsewhere using variogram models. The presented tomography-based method allows more control between wells and provides a realization-based way of assessing the positioning of reservoir boundaries. Figure \ref{fig:figure9} illustrates key milestones in the workflow which breaks down as follows: 
\begin{itemize}
\item
Estimating the tomography maximum-likelihood velocity model and computing target horizon error bars (resolved space) tuned to observed mis-ties at some well locations (Figure \ref{fig:figure9}a), 
\item
For the Top and Base reservoir horizons, describing the channel system of interest, calibrating horizon realizations to well markers (Figure \ref{fig:figure9}b). By doing so, uncertainties between wells are reflected by the spatial variations of horizon depth error bars derived from the tomographic operator.
\item
Estimating spill point depending on closure assumption for various horizon realizations.
Figure \ref{fig:figure9}b shows three fault-driven types of closure: ``four-way" closure made of dip or channel limits,``three-way" and ``two-way" closures add one or two sealed bounding faults. 
\item
Calculating GRV from Base and Top reservoir down to spill point closure level, with all these elements being affected by the error bars in the migrated domain and along well paths.
\item
Assessing, for each identified prospect, the closure probability map made from all plausible horizon realizations and defining the chance of finding closure above the measured spill point. Figure \ref{fig:figure9}c shows the probability map for one prospect, demonstrating the presence of robust structural closures in the same channel system.
\end{itemize}
This example emphasizes the importance of structural uncertainties for providing error bars between well locations
and allowing the generation of corresponding horizon realizations. 


\begin{figure}[H]
\centering
\begin{subfigure}
\centering
\includegraphics[width=1.\linewidth]{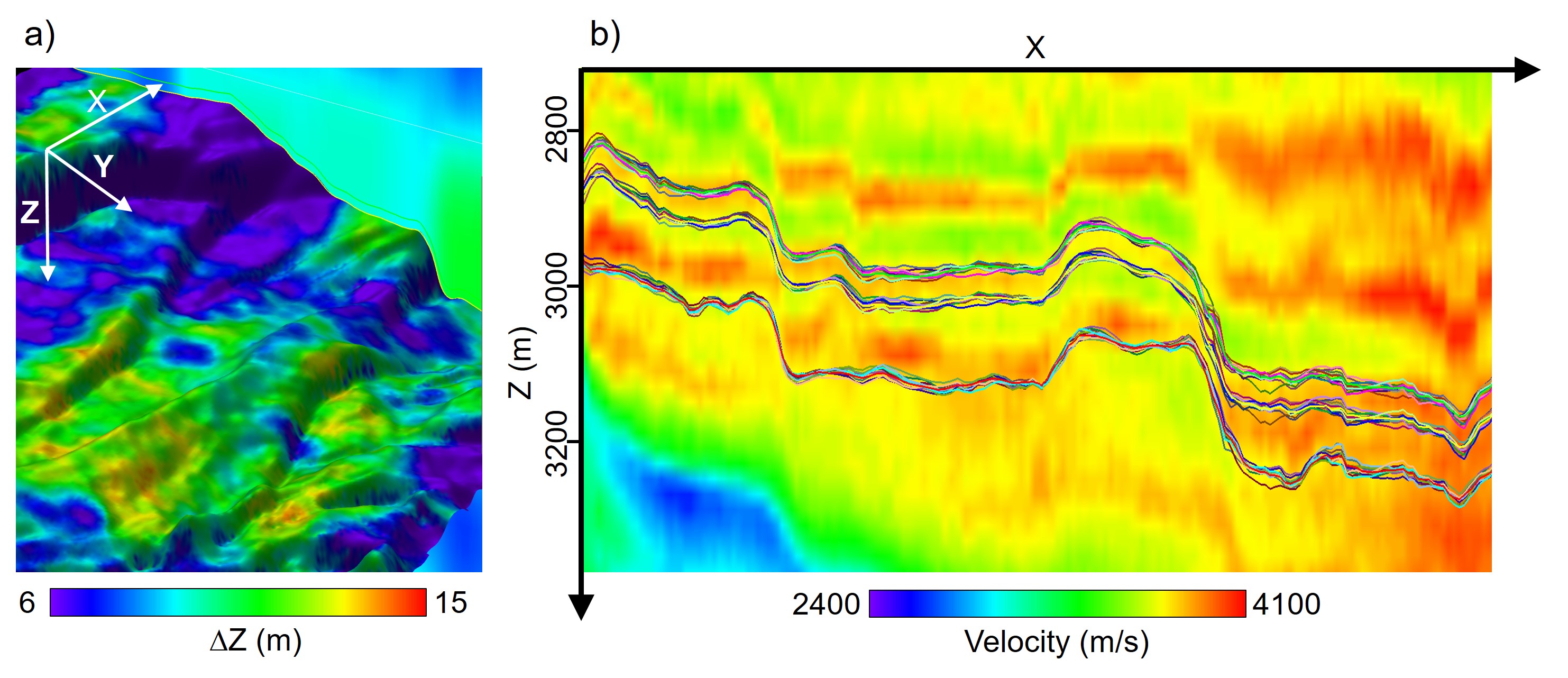}
\captionsetup{labelformat=empty}
\caption{
(a) Spatial variations of horizon depth error bars provide an assessment of the potential area with higher uncertainty. (b) Display of realizations of some key horizons superimposed on estimated velocity parameters.
}
\end{subfigure}
\begin{subfigure}
\centering
\includegraphics[width=1\linewidth]{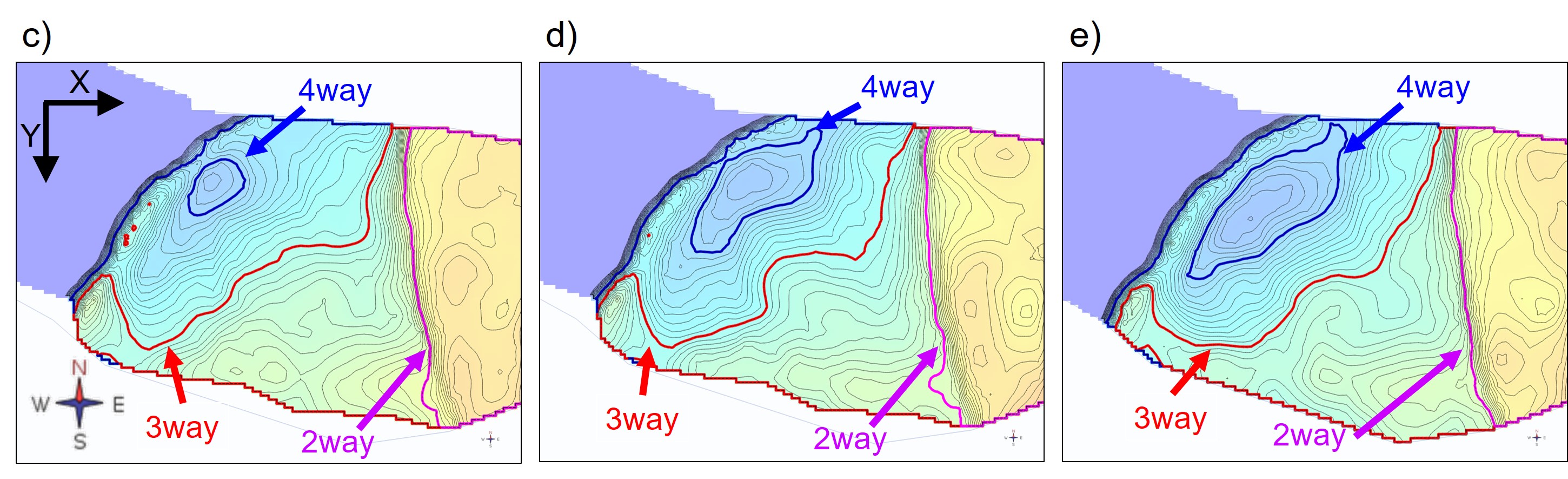}
\captionsetup{labelformat=empty}
\caption{
(c), (d), (e) Examples of reservoir contours for different spill point closure definitions and for different GRV scenarios (minimum case scenario on the left, average case in the center and maximum case on the right for the four-way closure type). For each scenario, the blue isoline represents the four-way closure, the red represents the three-way and the pink represents the two-way closure.
}
\end{subfigure}
\end{figure}
\begin{figure}[H]
\addtocounter{figure}{-3}
\begin{subfigure}
\centering
\includegraphics[width=1.\linewidth]{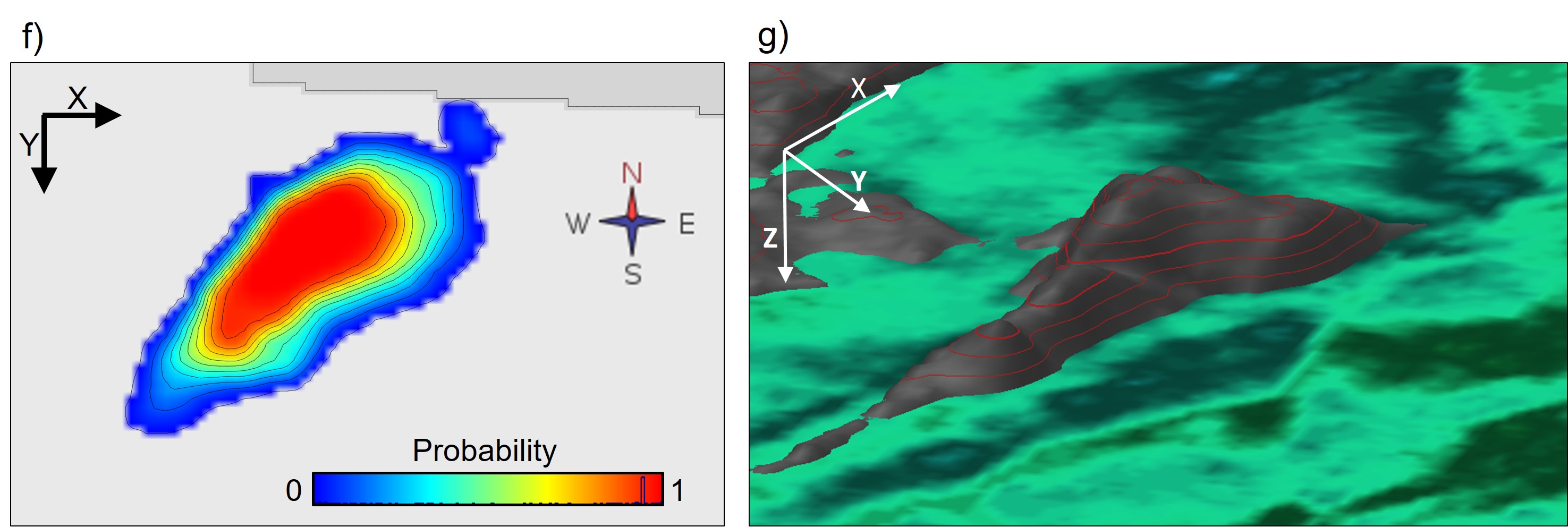}
\captionsetup{labelformat=empty}
\caption{
(f) Probability map of finding closure above the spill point for multiple realizations. (g) One realization of a calibrated top reservoir surface. 
}
\end{subfigure}
\caption{
Steps in GRV computation workflow and intermediate products or QCs.
Modified from \cite{Coleou2019_EAGE}; see the article for more details.
}
\label{fig:figure9}
\end{figure}

\section*{Conclusion}
\label{sec:concl}

We proposed an extension of previous industrial works
on the computation of migration structural uncertainties
and provided a unifying framework. 
Firstly, we estimated error bars from the statistical analysis of perturbed models obtained from the sampling of an equi-probable contour of the posterior pdf (related to a clear confidence level), not of the full pdf like in previous works.
Secondly, we developed the application in the context of non-linear slope tomography, based on the inversion of invariant picks. 
In addition to the advantages in terms of accuracy and efficiency of the VMB (compared to standard tomography)
it provides the possibility to assess the quality of the linear and Gaussian assumptions. 
It also allows us to compute volumetric migration positioning error bars
(using kinematic migration of all invariant picks and not only map migration of horizons invariant picks). 
Thirdly, we proposed to work in the full model space, not in a preconditioned model space (with smaller dimensionality). Splitting the analysis into the resolved and unresolved tomography spaces,
we argued that the resolved space uncertainties are to be used in further steps leading to decision-making and can be related to the output of methods that work in a preconditioned model space.
The unresolved space uncertainties represent a qualitative byproduct of our method
that reflects priors and the illumination,
strongly highlighting the most uncertain gross areas.
The latter can be useful for QCs.
These concepts were illustrated on a synthetic dataset.
Complementarily, the industrial viability of the method was illustrated on two field 3D datasets where emphasis was placed on the importance of horizon error bars,
among others for GRV computation and uncertainty evaluation between well locations. 

The presented approach can be applied to various E\&P topics.
Also, the approach can easily apply to FWI-derived models.
Indeed, state-of-the-art workflows involve interleaved FWI and tomography passes, ending with a tomography pass.
The corresponding tomography uncertainty analysis can naturally be performed to produce an estimate for FWI model kinematic-related uncertainties.


\section*{Acknowledgments}

The authors are particularly indebted to Mathieu Reinier, Herv\'e Prigent, Thierry Col\'eou, Jean-Luc Formento, Samuel Gray and Sylvain Masclet for collaboration and enlightening discussions.
The authors are grateful to CGG for granting permission to publish this work.
The authors thank Premier Oil, Edison, Bayerngas, WesternGeco, INEOS, OMV and CGG Multi-Client \& New Ventures Division for their permission to show the data. 


\newpage


\append{The binomial inverse theorem and the Woodbury matrix identity}

\label{App:0}


Consider an invertible (thus square) matrix $\mathbf{A}$ of size $N\times N$.
Consider a matrix $\mathbf{V}$ of size $N\times K_1$,
a matrix $\mathbf{\Lambda}$ of size $K_1 \times K_2$,
and a matrix $\mathbf{W}$ of size $K_2\times N$.
If the square matrix $\mathbf{I}_{K_1}+\mathbf{\Lambda}\mathbf{W}\mathbf{A}^{-1}\mathbf{V}$ of size $K_1 \times K_1$ is invertible,
we have the following identity, called binomial inverse theorem
\begin{eqnarray}
\big(
\mathbf{A}+\mathbf{V}\mathbf{\Lambda}\mathbf{W}
\big)^{-1}
=
\mathbf{A}^{-1}
-
\mathbf{A}^{-1}
\mathbf{V}
\big(
\mathbf{I}_{K_1}+\mathbf{\Lambda}\mathbf{W}\mathbf{A}^{-1}\mathbf{V}
\big)^{-1}
\mathbf{\Lambda}\mathbf{W}
\mathbf{A}^{-1}
.
\label{eq:Bin}
\end{eqnarray}
If the matrix $\mathbf{\Lambda}$ is invertible (thus square $K_1=K_2=K$),
the previous expression can be reduced to the Woodbury matrix identity
\begin{eqnarray}
\big(
\mathbf{A}+\mathbf{V}\mathbf{\Lambda}\mathbf{W}
\big)^{-1}
=
\mathbf{A}^{-1}
-
\mathbf{A}^{-1}
\mathbf{V}
\big(
\mathbf{\Lambda}^{-1}+\mathbf{W}\mathbf{A}^{-1}\mathbf{V}
\big)^{-1}
\mathbf{W}
\mathbf{A}^{-1}
.
\label{eq:Wood}
\end{eqnarray}

Consider $\mathbf{A}=\mathbf{I}_N$ ,
$\mathbf{\Lambda}$ square but not necessarilly invertible,
and $\mathbf{W}=\mathbf{V}^\dagger$ that satisfies $\mathbf{V}^\dagger\mathbf{V}=\mathbf{I}_{K}$.
If the matrix $\mathbf{I}_{K}+\mathbf{\Lambda}$ is invertible,
the binomial inverse theorem gives
\begin{eqnarray}
\big(
\mathbf{I}_N+\mathbf{V}\mathbf{\Lambda}\mathbf{V}^\dagger
\big)^{-1}
=
\mathbf{I}_N
-
\mathbf{V}
\big(
\mathbf{I}_{K}+\mathbf{\Lambda}
\big)^{-1}
\mathbf{\Lambda}\mathbf{V}^\dagger
.
\label{eq:Bin_2}
\end{eqnarray}
Using the binomial theorem again under same conditions,
we obtain
$
(\mathbf{I}_{K}+\mathbf{\Lambda})^{-1}\mathbf{\Lambda}
=
\mathbf{I}_{K}-(\mathbf{I}_{K}+\mathbf{\Lambda})^{-1}
$),
and finally the following identity
\begin{eqnarray}
\big(
\mathbf{I}_N+\mathbf{V}\mathbf{\Lambda}\mathbf{V}^\dagger
\big)^{-1}
=
\mathbf{I}_N
-
\mathbf{V}
\mathbf{V}^\dagger
+
\mathbf{V}
\big(
\mathbf{I}_{K}+\mathbf{\Lambda}
\big)^{-1}
\mathbf{V}^\dagger
.
\label{eq:Bin_3}
\end{eqnarray}

\append{Standard-deviations and error bars}
\label{App:-2}


The multi-dimensional Gaussian posterior pdf, equation \ref{eq:post_pdf}, can be rewritten
(we do not consider the normalization factor here to simplify the notations
without loss of generality,
so that the maximum of the pdf is always $1$)
\begin{eqnarray}
\tilde\rho_M(\Delta\mathbf{m})
&= &
\exp\big[
-\frac{1}{2}
\Delta\mathbf{m}^\dagger
\tilde{\mathbf{C}}_M^{-1}
\Delta\mathbf{m}
\big]
\label{eq:post_pdf_app}\\
&= &
\prod_{i=1}^{N_M}
\tilde\rho_{M_i}(\Delta\mathbf{m})
,
\nonumber
\end{eqnarray}
where the posterior pdf for one model space node $i$ is defined by
\begin{eqnarray}
\tilde\rho_{M_i}(\Delta\mathbf{m})
&= &
\exp\big[
-\frac{1}{2}
\tilde{{C}}_{M_{ii}}^{-1}
\Delta{m}_i^2
\big]
\exp\big[
-\frac{1}{2}
A_i(\Delta\mathbf{m})
\Delta{m}_i
\big]
\nonumber\\
A_i(\Delta\mathbf{m})
&=&
\sum_{\substack{j=1\\j\ne i}}^{N_M}
\tilde{{C}}_{M_{ij}}^{-1}\Delta{m}_j
.
\label{eq:post_pdf_i2}
\end{eqnarray}
%
$A_i(\Delta\mathbf{m})\Delta{m}_i$ may be negative
and can be neglected only when the correlations $\tilde{{C}}_{M_{i,j\ne i}}^{-1}$ are sufficiently small.
%

Consider the models related to a given value $a\in[0,1]$
of the un-normalized posterior pdf for one model space node, i.e.
$
\forall i=1..N_M:
\tilde\rho_{M_i}(\Delta\mathbf{m})=a
$.
In other terms,
$a$ denotes a percentage of the maximum of each $\tilde\rho_{M_i}$.
Using equation \ref{eq:post_pdf_app}, this leads to
\begin{eqnarray}
&&
\tilde\rho_M(\Delta\mathbf{m}) = a^{N_M}
,
\nonumber\\
&&
\Delta\mathbf{m}^\dagger
\tilde{\mathbf{C}}_M^{-1}
\Delta\mathbf{m}
=
2\ln(1/a)\times N_M
.
\label{eq:post_pdf_app4}
\end{eqnarray}
This equation defines the set of model perturbations $\Delta\mathbf{m}$
related to a chosen $a$ value for the
posterior pdfs of each model space node $\tilde\rho_{M_i}(\Delta\mathbf{m})$.
%
Equation \ref{eq:post_pdf_app4} is equivalent to
\begin{eqnarray}
\sum_{i=1}^{N_M}
\tilde{{C}}_{M_{ii}}^{-1}
\Delta{m}_i^2
+
\sum_{i=1}^{N_M}
A_i(\Delta\mathbf{m})
\Delta m_i
=
2\ln(1/a)\times N_M
.
\label{eq:post_pdf_app2}
\end{eqnarray}
In the general case, one has to resolve the full equation \ref{eq:post_pdf_app2}
to obtain the solutions $\Delta{m}_i$.
But if $|\sum_{i=1}^{N_M}
A_i(\Delta\mathbf{m})
\Delta m_i|
<<
|\sum_{i=1}^{N_M}
\tilde{{C}}_{M_{ii}}^{-1}
\Delta{m}_i^2|
$,
the solutions become
\begin{eqnarray}
\Delta{m}_i
\quad
=
\pm
\tilde{{C}}_{M_{ii}}^{1/2}
\sqrt{2\ln(1/a)}
,
\label{eq:post_pdf_app6}
\end{eqnarray}
i.e. error bars are then related to the posterior standard deviations $\tilde{{C}}_{M_{ii}}^{1/2}$.
Error bars become equal to $\pm \tilde{{C}}_{M_{ii}}^{1/2}$
if we select $a=0.6$,
that is related to a confidence probability $P(a)=68.3\%$
(computing the correctly normalized integral of $\tilde\rho_M(\Delta\mathbf{m})$ inside the corresponding hyper-ellipsoid).


\newpage

\bibliographystyle{seg} 
\bibliography{biblio}

\begin{thebibliography}{}
\itemsep0pt

\bibitem[Adler et~al., 2008]{Adler2008}
Adler, F., R. Baina, M.~A. Soudani, P. Cardon, and J.-B. Richard,  2008,
  Nonlinear 3d tomographic least-squares inversion of residual moveout in
  kirchhoff prestack-depth-migration common image gathers: Geophysics, {\bf
  73}, VE2--VE13.

\bibitem[Al~Chalabi, 1994]{AlChalabi1994}
Al~Chalabi, M.,  1994, Seismic velocities, a critique: First Break, {\bf 12},
  589--596.

\bibitem[Allemand et~al., 2017]{Allemand2017}
Allemand, T., A. Sedova, and O. Hermant,  2017, Flattening common image gathers
  after full-waveform inversion: the challenge of anisotropy estimation: 87th
  SEG Annual Meeting, Expanded Abstracts,  1410--1415.

\bibitem[Boor, 1978]{DeBoor1978}
Boor, D.,  1978, A practical guide to splines: Springer Verlag.

\bibitem[Chauris et~al., 2002]{Chauris2002}
Chauris, H., M. Noble, G. Lambar\'s, and P. Podvin,  2002, Migration velocity
  analysis from locally coherent events in 2-d laterally heterogeneous media,
  part i : Theoretical aspects: Geophysics, {\bf 67}, 1213--1224.

\bibitem[Choi, 2006]{Choi06}
Choi, S.-C.,  2006, Iterative methods for singular linear equations and
  least-squares problems: PhD thesis, Stanford University.

\bibitem[Clapp et~al., 1998]{Cla98}
Clapp, R.~G., B.~L. Biondi, S. Fomel, and J.~F. Claerbout,  1998, Regularizing
  velocity estimation using geologic dip information: SEG, Expanded Abstracts.

\bibitem[Col\'eou et~al., 2019]{Coleou2019_EAGE}
Col\'eou, T., J.-L. Formento, H. Prigent, J. Messud, D. Laurencin, M. Reinier,
  P. Guillaume, A. Egreteau, and L. Damian,  2019, Use of tomography velocity
  uncertainty in grv calculation.: 81st EAGE Conference and Exhibition,
  Workshop.

\bibitem[Cowan, 1998]{Cow98}
Cowan, G.,  1998, Statistical data analysis: Oxford Science Publications.

\bibitem[Duffet and Sinoquet, 2002]{Duffet2002}
Duffet, C., and D. Sinoquet,  2002, Quantifying geological uncertainties in
  velocity model building: 72nd SEG Annual Meeting, Expanded Abstracts,
  926--929.

\bibitem[Duffet and Sinoquet, 2006]{Duf06}
--------, 2006, Quantifying uncertainties on the solution model of seismic
  tomography: Inverse Problems, {\bf 22}, 525--538.

\bibitem[Fournier et~al., 2013]{Fou13}
Fournier, A., N. Ivanova, Y. Yang, K. Osypov, C. Yarman, D. Nichols, R.
  Bachrach, Y. You, M. Woodward, and S. Centanni,  2013, Quantifying e\&p value
  of geophysical information using seismic uncertainty analysis: 75th EAGE
  Conference and Exhibition.

\bibitem[Guillaume et~al., 2008]{Guillaume2008}
Guillaume, P., G. Lambar\'e, O. Leblanc, P. Mitouard, J. Le~Moigne, J. Montel,
  A. Prescott, R. Siliqi, N. Vidal, X. Zhang, and S. Zimine,  2008, Kinematic
  invariants: an efficient and flexible approach for velocity model building:
  78th SEG Annual Meeting, Expanded Abstracts,  3687--3692.

\bibitem[Guillaume et~al., 2011]{Guillaume2011}
Guillaume, P., G. Lambar\'e, S. Sioni, D. Carotti, P. D\'epr\'e, G. Culianez,
  J.-P. Montel, P. Mitouard, S. Depagne, S. Frehers, and H. Vosberg,  2011,
  Geologically consistent velocities obtained by high definition tomography:
  81st SEG Annual Meeting, Expanded Abstracts,  4061--4065.

\bibitem[Guillaume et~al., 2013a]{Guillaume2013b}
Guillaume, P., M. Reinier, G. Lambar\'e, A. Cavali\'e, M. Adamsen, and B.
  Bruun,  2013a, Dip constrained non-linear slope tomography - an application
  to shallow channel characterization: 75th EAGE Conference and Exhibition.

\bibitem[Guillaume et~al., 2013b]{Guillaume2013}
Guillaume, P., X. Zhang, A. Prescott, G. Lambar\'e, M. Reinier, J.-P. Montel,
  and A. Cavali\'e,  2013b, Multi-layer non-linear slope tomography: 75th EAGE
  Conference and Exhibition.

\bibitem[Hajnal and Sereda, 1981]{Hajnal1981}
Hajnal, Z., and I.~T. Sereda,  1981, Maximum uncertainty of interval velocity
  estimates: Geophysics, {\bf 46}, 1543--1547.

\bibitem[Lambar\'e, 2008]{Lambare2008}
Lambar\'e, G.,  2008, Stereotomography: Geophysics, {\bf 73}, VE25--VE34.

\bibitem[Lambar\'e et~al., 2014]{Lambare2014}
Lambar\'e, G., P. Guillaume, and J.-P. Montel,  2014, Recent advances in
  ray-based tomography: 76th EAGE Conference and Exhibition.

\bibitem[Li et~al., 2014]{Li14}
Li, L., J. Caers, and P. Sava,  2014, Uncertainty maps for seismic images
  through geostatistical model randomization: SEG, Expanded Abstracts,
  1496--1500.

\bibitem[Messud et~al., 2017a]{Messud2017_EAGE}
Messud, J., P. Guillaume, and G. Lambar\'e,  2017a, Estimating structural
  uncertainties in seismic images using equi-probable tomographic model: 79th
  EAGE Conference and Exhibition.

\bibitem[Messud et~al., 2018]{Messud2018_EAGE}
Messud, J., P. Guillaume, M. Reinier, and C. Hidalgo,  2018, Migration
  confidence analysis: Resolved space uncertainties: 80th EAGE Conference and
  Exhibition.

\bibitem[Messud et~al., 2017b]{Messud2017_TLE}
Messud, J., M. Reinier, H. Prigent, P. Guillaume, T. Col\'eou, and S. Masclet,
  February 2017b, Extracting seismic uncertainties from tomographic velocity
  inversion and their use in reservoir risk analysis: The Leading Edge, SEG,
  {\bf 36}, 127--132.

\bibitem[Mun\~oz and Rath, 2006]{Mun06}
Mun\~oz, G., and V. Rath,  2006, Beyond smooth inversion: the use of nullspace
  projection for the exploration of non-uniqueness in mt: Geophys. J. Int.,
  {\bf 164}, 301--311.

\bibitem[Operto et~al., 2003]{Operto2003}
Operto, S., G. Lambar\'e, P. Podvin, and P. Thierry,  2003, 3d ray+born
  migration/inversion-part2: Application to the seg/eage overthrust experiment:
  Geophysics, {\bf 68}, 1357--1370.

\bibitem[Osypov et~al., 2008a]{Osy08a}
Osypov, K., D. Nichols, M. Woodward, and C.~E. Yarman,  2008a, Tomographic
  velocity model building using iterative eigendecomposition: 70th EAGE
  Conference and Exhibition.

\bibitem[Osypov et~al., 2008b]{Osy08b}
Osypov, K., D. Nichols, M. Woodward, O. Zdraveva, and C.~E. Yarman,  2008b,
  Uncertainty and resolution analysis for anisotropic tomography using
  iterative eigendecomposition: SEG, Expanded Abstracts,  3244--3249.

\bibitem[Osypov et~al., 2010]{Osypov2010}
Osypov, K., D. Nichols, Y. Yang, F. Qiao, M. O'Briain, and O. Zdraveva,  2010,
  Quantifying structural uncertainty in anisotropic depth-imaging. gulf of
  mexico case study: 72nd EAGE Conference and Exhibition.

\bibitem[Osypov et~al., 2011]{Osy11}
Osypov, K., M. O'Briain, P. Whitfield, D. Nichols, A. Douillard, P. Sexton, and
  P. Jousselin,  2011, Quantifying structural uncertainty in anisotropic model
  building and depth imaging: Hild case study: 73rd EAGE Conference and
  Exhibition.

\bibitem[Osypov et~al., 2013]{Osypov2013}
Osypov, K., Y. Yang, A. Fourier, N. Ivanova, R. Bachrach, C.~E. Yarman, Y. You,
  D. Nichols, and M. Woodward,  2013, Model-uncertainty quantification in
  seismic tomography: method and applications: Geophysical Prospecting, {\bf
  61}, 1114--1134.

\bibitem[Paige and Saunders, 1982]{Pai82}
Paige, C.~C., and M.~A. Saunders,  1982, Lsqr: An algorithm for sparse linear
  equations and sparse least squares: ACM Transactions on Mathematical
  Software, {\bf 8}, 43--71.

\bibitem[Reinier et~al., 2017]{Reinier2017_EAGE}
Reinier, M., J. Messud, P. Guillaume, and T. Rebert,  2017, Tomographic model
  uncertainties and their effect on imaged structures: 79th EAGE Conference and
  Exhibition, Workshop.

\bibitem[Simpson et~al., 2000]{Simpson2000}
Simpson, G., F. Lamb, J. Finch, and N. Dinnie,  2000, The application of
  probabilistic and qualitative methods to asset management and decision
  making: SPE.

\bibitem[Tarantola, 1986]{Tarantola1986}
Tarantola, A.,  1986, A strategy for nonlinear elastic inversion of seismic
  reflection data: Geophysics, {\bf 51}, 1893--1903.

\bibitem[Tarantola, 2005]{Tar05}
--------, 2005, Inverse problem theory and methods for model parameters
  estimation: Society for Industrial and Applied Mathematics (Philadelphia).

\bibitem[Thore et~al., 2002]{Thore2002}
Thore, P., A. Shtuka, M. Lecour, T. Ait-Ettajer, and R. Cognot,  2002,
  Structural uncertainties: Determination, management, and applications:
  Geophysics, {\bf 67}, 840--852.

\bibitem[Virieux and Farra, 1991]{Virieux1991}
Virieux, J., and V. Farra,  1991, Ray tracing in 3-d complex isotropic media:
  An analysis of the problem: Geophysics, {\bf 56}, 2057--2069.

\bibitem[Virieux and Operto, 2009]{Virieux2009}
Virieux, J., and S. Operto,  2009, An overview of full-waveform inversion in
  exploration geophysics: Geophysics, {\bf 74}, WCC1--WCC26.

\bibitem[Woodward et~al., 2008]{Woodward2008}
Woodward, M., D. Nichols, O. Zdraveva, P. Whitfield, and T. Johns,  2008, A
  decade of tomography: Geophysics, {\bf 73}, VE5--VE11.

\bibitem[Zhang and Thurber, 2007]{Zha07}
Zhang, H., and C.~H. Thurber,  2007, Estimating the model resolution matrix for
  large seismic tomography problems based on lanczos bidiagonalization with
  partial reorthogonalization: Geophys. J. Int., {\bf 170}, 337--345.

\bibitem[Zhang and McMechan, 1995]{Zhang1995}
Zhang, J., and G.~A. McMechan,  1995, Estimation of resolution and covariance
  for large matrix inversions: Geophys. J. Int., {\bf 121}, 409--426.

\end{thebibliography}

\end{document}